\documentclass[]{article}

\usepackage{amssymb,latexsym,cite}
\usepackage[latin1]{inputenc}
\usepackage{amsmath}

\usepackage[dvips]{color}
\usepackage{dsfont}
\usepackage[OT1]{fontenc}
\usepackage{psfrag,float}
\usepackage{graphicx,subfig}

\newcommand{\Real}{\mathbb{R}}
\newcommand{\Integer}{\mathbb{Z}}
\newcommand{\Complex}{\mathbb{C}}
\newcommand{\Natural}{\mathbb{N}}

\newcommand{\RealPart}{Re}

\newcommand{\sign}[1]{{{\text{sign } #1}}}

\newtheorem{theorem}{Theorem}[section]
\newtheorem{lemma}[theorem]{Lemma}

\newtheorem{corollary}[theorem]{Corollary}

\newcommand{\qed}{\nobreak \ifvmode \relax \else
  \ifdim\lastskip<1.5em \hskip-\lastskip
  \hskip1.5em plus0em minus0.5em \fi \nobreak
  \vrule height0.75em width0.5em depth0.25em\fi}

\begin{document}
\title{Finite Difference Methods for 
  Second Order in Space, First Order in Time Hyperbolic Systems 
  and the Linear Shifted Wave Equation as a Model Problem in Numerical Relativity}  
  \author{M. Chirvasa\footnote{Max-Planck-Institut f\"ur Gravitationsphysik, Am M\"uhlenberg 1, 14475 Potsdam, Germany.} and S. Husa\footnote{Departament de F\'{\i}sica, Universitat de les Illes
      Balears, Cra.~Valldemossa Km.~7.5, Palma de Mallorca, E-07122 Spain.}}

  \maketitle
  \begin{abstract}
   Motivated by the problem of solving the Einstein equations,
   we discuss high order finite difference discretizations of
   first order in time, second order in space hyperbolic systems. 
   Particular attention is paid to the case when first order
   derivatives that can be identified with advection terms are approximated
   with non-centered finite difference operators.
   We first derive general properties of these discrete operators,
   then we extend a known result on numerical stability for 
   such systems to general order of accuracy.
   As an application we analyze the shifted wave equation,
   including the behavior of the numerical phase and group speeds
   at different orders of approximations.
   Special attention is paid to 
   when the use of off-centered schemes improves the accuracy 
   over the centered schemes.
  \end{abstract}
  
  \section {Introduction}
  Numerical discretization of first order hyperbolic systems of 
  partial differential equations (PDEs) 
  is greatly simplified by a result
  for the linear constant coefficient case
  \cite{Kreiss}: If the Cauchy problem is well-posed,
  then the semi-discrete problem (only discretizing space and leaving
  time continuous) is stable when spatial derivatives are discretized
  with a centered finite difference operator (CFDO).
  Furthermore, when using simple Runge-Kutta methods \cite{Hair} 
  for time integration, for sufficiently small time step
  the fully discrete problem is also stable.

  Such a result does {\em not} hold in general for second order systems where
  first {\em and} second spatial derivatives appear \cite{CalHinHusa}! 
  In order to obtain a stable semi-discrete
  scheme, the second order system needs to have additional properties.
  In \cite{CalHinHusa}, which in the following we refer to as CHH,
  sufficient conditions for stability of the fully discrete
  problem were presented for such systems. 
  Although these conditions are valid for general order centered 
  discretizations, this point has not been explicitly made. 
  One of the results of this  article is to make this statement clear, 
  by closing a technical gap related to the boundedness of the lower order terms.
  The main focus of our work is on a detailed analysis of the numerical
  properties  of discretizations  where 
  some first order derivatives are approximated with off-centered 
  finite difference operators and artificial dissipation is added to
  the equations.
  The motivation for choosing this more general situation comes from 
  numerical relativity, where it is a common practice to off-center by 
  one point the derivatives corresponding to the Lie advection terms. 
  In numerical simulations of black holes using the BSSN formulation
  of the Einstein equations 
  \cite{Shibata95,Baumgarte99,Gundlach:2006tw}, 
  this procedure of off-centering was found
  to be essential even for sixth order schemes \cite{Husa:2007hp}.
  Numerical solutions of the  Einstein  equations are currently
  quickly expanding our knowledge about the astrophysics of compact
  binaries (see \cite{Pretorius:2007nq,Hannam:2009rd} for overviews
  on what has been achieved since the  major breakthroughs in 2005
  \cite{Pretorius:2005gq,Campanelli:2005dd,Baker:2005vv}), but a systematic
  understanding of the underlying  numerical techniques  has  not yet
  been achieved.

  The shifted scalar wave equation serves as a simple but 
  powerful model in numerical relativity 
  \cite{Szi,Babiuc:2004pi,Babiuc:2005fr,GundCal,HarmCode}. 
  The  particular case with zero shift and flat background 
  (standard wave equation) has been extensively studied 
  in \cite{Cohen}-\cite{Anne} and high order 
  discretization methods have been proposed.
  In Sec. \ref{sec:Prelim} we introduce the 
  shifted scalar wave equation as 
  a first order in time, second order in space system, 
  together with a summary of the well-posedness theory 
  for mixed order systems.
  We show stability for our semi-discrete problem, independent of
  shift or dissipation terms, while the  fully-discrete problem 
  requires artificial dissipation if more than one point is off-centered.
  Restricting to flat space in one space dimension, Courant limits 
  and numerical phase and group speeds are computed 
  and analyzed in detail. It is shown that increasing the off-centering reduces 
  the Courant limit. However, by increasing the order of approximation 
  while keeping the off-centering fixed, does not necessarily generate 
  lower Courant limits.
  Regarding the numerical speeds, it is shown that indeed there are cases  
  when off-centering improves the accuracy over the centered scheme.
  This fact is illustrated also experimentally by the results of some 
  simple numerical tests at the end of Section \ref{sectionWave}. 
  
  Our analysis of the wave equation relies on certain properties of 
  finite difference operators, in particular on their behavior in Fourier space.
  We introduce these operators in Section \ref{sec:FDO} 
  together with highlighting some relevant properties. 
  Then, in Section \ref{sec:StabMethod} 
  we address the stability method and follow in Section \ref{sectionWave} 
  with the analysis of the wave equation. 
  Our results are summarized in Section \ref{sec:summary}.

\section{The shifted wave equation and
     first order in time
        second order in space hyperbolic systems}
\label{sec:Prelim}

The scalar wave equation in a $d+1$-dimensional spacetime equipped with a 
Lorentzian metric $ g_{\alpha\beta}$ reads
\begin{equation}
  g^{\alpha\beta}\partial_{\alpha}\partial_{\beta}\mathsf{\Phi}=0.
\label{gwaveeq}
\end{equation}
We assume a uniform
time slicing for simplicity, $g^{00}=-1$, and perform a $d+1$ split 
introducing a positive definite 
d-metric $\gamma^{jl}=g^{jl}+\beta^{i}\beta^{j}$,
with $i,j=\overline{1,d}$ and a shift vector $\beta^{i}=g^{0i}$ (see e.g. \cite{Wald}).
The wave equation (\ref{gwaveeq}) then becomes
\begin{equation}
  \partial_{tt}\mathsf{\Phi}=2\beta^{i}\partial_{i}\partial_{t}\mathsf{\Phi}
  +\left(\gamma^{jl}-\beta^{i}\beta^{j}\right)\partial_{i}\partial_{j}\mathsf{\Phi}\,.
\nonumber
\end{equation}
The mixed time-space derivatives lead to non-standard behavior 
as compared to the flat space wave equation with zero shift, 
and much of the material below will be devoted to their treatment.

We reduce the wave equation to a first order in time, second order
in space form by introducing the variable $\mathsf{K}$, 
in analogy with the York-ADM-system \cite{York_in_Smarr} (and other common
representations of the Einstein equations),
\begin{equation}
  \mathsf{K}=\partial_{t}\mathsf{\Phi}-\beta^{j}\partial_{j}\mathsf{\Phi}
\nonumber\end{equation}
which transforms the wave equation into the first order in time, second order
in space system in the way most common in numerical relativity:
\begin{equation}
  \partial_{t}\mathsf{\Phi}=\beta^{j}\partial_{j}\mathsf{\Phi}+\mathsf{K}\,,\qquad
  \partial_{t} \mathsf{K}=\gamma^{jl}\partial_{jl}\mathsf{\Phi}+\beta^{j}\partial_{j}\mathsf{K}\,.
  \label{contwaveeq3D}
\end{equation}
Well-posedness for the Cauchy problem 
for the system (\ref{gwaveeq}) is a standard textbook result both in the original
second order form and for reduction to first order symmetric hyperbolic form. 
In the latter form standard theorems for numerical stability apply \cite{Kreiss}.
Here we investigate the numerical stability for the first order in time, 
second order in space system (\ref{contwaveeq3D}),  
using the methods presented in \cite{CalHinHusa}.
In this respect, the  appropriate generalization
of the  shifted wave equation is a linear system of PDEs with constant 
coefficients  of the form  \cite{CalHinHusa}: 
\begin{equation}
  \frac{d}{dt}\mathsf{v}(t,x)=\mathsf{P} \mathsf{v}(t,x),\,\,\,
  \mathsf{v}=(\mathsf{U},\mathsf{V})^{T} \,,
  \nonumber\end{equation}  
  with $ x \in\mathbb{R}^{d}$,  
  $ \mathsf{U}:\mathbb{R}\text{x}\mathbb{R}^{d}\rightarrow\mathbb{R}^{p}$,
  $ \mathsf{V}:\mathbb{R}\text{x}\mathbb{R}^{d}\rightarrow\mathbb{R}^{q}$
  and 
    \begin{eqnarray}  
      \mathsf{P}&=&\left(
      \begin{array}{cc}
        A^{j}\partial_{j}+B&C\\
        D^{jl}\partial_{jl}+E^{j}\partial_{j}+F&G^{j}\partial_{j}+J
      \end{array}
      \right)\,.
      \label{contsystem}
    \end{eqnarray}
    Note that the state vector $v$ is split into
    two parts, $U$ are those variables for which only
    first spatial derivatives appear, while
    second spatial derivatives of the $V$-variables
    do enter the PDE.
    The well-posedness of the Cauchy problem for 
    first order in time, second order in space systems of PDEs
    systems has been clarified by \cite{SCPT,NOR,GG1,GG2,BS}.
    We will here recall the presentation 
    in CHH, where the well-posedness of such systems of PDEs is discussed
    in close analogy with the issue of numerical stability.  
    It is natural to consider $2\pi$-periodic solutions 
    and turn the analysis in Fourier space.
  
    In Fourier space the evolution 
    problem reduces to a system of ordinary differential equations (ODEs) 
    for the Fourier coefficients.
    By performing a first order reduction in Fourier space, 
    it can be shown that well-posedness is not influenced 
    by lower differential order terms, which we can therefore drop
    and consider  
    the second order principal symbol constructed as
     \begin{equation}
       \hat{\mathsf{P}}'=\left(
       \begin{array}{cc}
	 i\omega_{0} A^{n}&C \\
	 {-}\omega_{0}^{2}D^{nn} &i\omega_{0} G^{n}
       \end{array}
       \right)\,,
       \nonumber\end{equation}
       where  $\omega_{0}=\left\vert\omega\right\vert$, 
       $\omega=(\omega_{1},\dots,\omega_{d}) \in \Integer^{d}$, 
       $M^{n}=M^{j}n_{j}$ and   $n_{j} =\omega_{j} \omega_{0}^{-1}$. 
      
     It is shown in \cite{CalHinHusa} that if there exists a matrix 
  $\hat{\mathsf{H}}(\omega)=\hat{\mathsf{H}}^{\dagger}(\omega)$ 
     such that
     $\hat{\mathsf{H}}\hat{\mathsf{P}}'+
     \hat{\mathsf{P}}'^{\dagger}\hat{\mathsf{H}}
     =0$
     and a positive constant $K$,
     such that $K^{-1}I_{\omega_{0}}\leq \hat{\mathsf{H}}\leq K I_{\omega_{0}}$
     (where $I_{\omega_{0}}=\text{diag}[\omega_{0}^{2}I_{p},I_{q}]$),
     then the problem is well-posed in the norms
     \begin{equation}
       \left\Vert \mathsf{v}\right\Vert^{2}_{\partial}=\int
       \sum_{j=1}^{d}
       \left\vert \partial_{j}\mathsf{U}\right\vert^{2}+\left\vert
       \mathsf{V}\right\vert^{2}\,,
 \qquad
    \left\Vert\mathsf{v}\right\Vert_{\mathsf{H}}^{2}
    =\sum_{\omega \in \Integer^{d}}
    \hat{\mathsf{v}}^{\dagger}\hat{\mathsf{H}}
    \hat{\mathsf{v}}\,.
  \nonumber\end{equation}

In \cite{CalHinHusa} an analysis of numerical stability  was  performed
in analogy with the proof of well-posedness as we have sketched it, and
which we will extend  to arbitrary approximation order in Section  
\ref{sec:StabMethod}. But, before going into stability analysis, 
we need to discuss some general properties of finite difference 
operators.

\section{Finite Difference Operators}\label{sec:FDO}
\subsection{Construction and properties in one dimension}\label{constr_prop_1D}
 Consider a mesh of equidistant points $x_{\nu}=\nu h$, 
 with $\nu\in\Integer$ and $h$ representing the grid spacing.
  Corresponding to the continuum vector function 
  $\mathsf{v}:\Real\rightarrow \Complex\times\dots\times \Complex$ 
  we associate the grid vector function $v$ by  
  $v:\{x_{\nu },\,\nu =0,\pm 1,\pm 2,\dots\}\rightarrow 
  \Complex\times\dots\times \Complex$ and 
  $v_{\nu }\equiv v(x_{\nu })=\mathsf{v}(x_{\nu })$.
  
  Using $2n+1$ consecutive points, 
  we want to construct the finite difference operator 
  corresponding to the $m$-derivative.
  Let  $s\in\{0,1,\dots, n \}$ be the offset of these 
  points from symmetry 
  with respect to the center, ($s=0$ for CFDO) and 
  $\epsilon$ the direction of off-centering
  ($\epsilon=1$ for off-centering to the right, $\epsilon=-1$ for off-centering
  to the left).\footnote{Though one can simplify the notation by dropping
  $\epsilon$ and considering $s\in\{-n,\dots,n\}$, 
    it will later turn out useful to separate the sign of 
  $s$ and its absolute value.} 
  Then the finite difference operator to be constructed will be denoted 
  $D^{(m,n,s,\epsilon)}$. It is a linear combination of shift operators 
  of the form:
  \begin{equation}
    D^{(m,n,s,\epsilon)}=
    h^{-m}\sum_{k=-n+\epsilon s}^{n+\epsilon s}\tilde{f}_{m,n,s,\epsilon,k}S^{k}\,,
    \label{genformFDOexplicit}
  \end{equation} 
   where  $S^{k}$ be the shift operator by $k$ points, $S^{k}v_{\nu }=v_{\nu +k}$.
   The weights $\tilde{f}_{m,n,s,\epsilon,k}$ can be expressed
   as the coefficients of $y^{k}$ in the Taylor expansion of the function 
   $$f^{m,n,s,\epsilon}(y)=y^{n-\epsilon s}(\ln y)^{m}$$ 
   around the point $y_{0}=1$ up to the order $(y-y_{0})^{2n}$ 
   (see appendix \ref{constr_prop_1D_explicit} for the proof).
   Using this procedure, one can deduce 
   explicit expressions for the finite difference 
   operators corresponding to the first and second derivative 
   (the relations (\ref{explicitFDO}) 
   from appendix \ref{constr_prop_1D_explicit}).

   These expressions are fairly complicated, 
   but they can be written 
   in a more convenient form if we make use of 
   the elementary finite difference operators:
   \begin{eqnarray}
     D_{\pm}v_{\nu}&=& \pm h^{-1}(v_{\nu\pm 1}-v_{\nu}),\nonumber\\
     \delta_{0} &=&\frac{h}{2}\left(D_{+}+D_{-}\right),\nonumber\\
     p&=&h(D_{+}-D_{-})=h^{2}D_{+}D_{-}\,.
     \label{elemoperators}
   \end{eqnarray}
   Note that the operators $\delta_{0}$ and $p$ 
   are dimensionless.

   Then, a direct but lengthy calculation 
   starting from the definitions (\ref{explicitFDO})  
   leads us to the following expressions for 
   $ D^{(1,n)}\equiv D^{(1,n,0,0)}$, 
   $ D^{(2,n)}\equiv D^{(2,n,0,0)}$, 
   the rest $R^{(n)}\equiv(D^{(1,n)})^{2}-D^{(2,n)}$ 
   and    $ D^{(1,n,s,\epsilon)}$:
   \begin{eqnarray}
     D^{(1,n)}&=&h^{-1}\delta_{0}\left(1+\sum_{k=1}^{n-1}c_{k}p^{k}\right)\,,
     \nonumber\\
     D^{(2,n)}&=&h^{-2}p\left(1+\sum_{k=1}^{n-1}d_{k}p^{k}\right)\,,
      \nonumber\\
      R^{(n)}&=&h^{-2}\frac{n c_{n-1}}{2}p^{n+1}
      \sum_{k=0}^{n-1}\frac{c_{k}}{n+1+k}p^{k}\,.
      \nonumber\\
      D^{(1,n,s,\epsilon)}&=&D^{(1,n)}+h^{-1}\left(
      \delta_{0}\sum_{k=1}^{s-1}a_{k}p^{k}
      -\epsilon
      p\sum_{k=1}^{s-1}b_{k}p^{k}
      \right)p^{n}\,.
    \label{FDOconvenient}
    \end{eqnarray}
    where 
    \begin{eqnarray}
      c_{k}=(-1)^{k}\frac{(k!)^{2}}{\left(2k+1\right)!},
      &&d_{k}=\frac{c_{k}}{k+1}\,,    \nonumber\\
      a_{k}=(-1)^{n}\sum_{j =k}^{s-1}
      \frac{(-1)^{j }C_{j +k}^{2k}}{(n-j )C_{2n}^{n+j }},
      &&b_{k}=(-1)^{n+s}\frac{C_{s+k}^{2k+1}}{2(n+1+k)C_{2n}^{n+s}}
    \end{eqnarray}
  Notice that the coefficients $c_k$ and $d_k$ 
  do not depend on $n$, while $a_{k}$, $b_{k}$ do depend 
  on $n$ and $s$.
  
  The leading order truncation error of order $2n$ is defined as
  \begin{eqnarray*}
    \left.\frac{dv}{dx}\right|_{x_{0}}-D^{(1,n,s,\epsilon)}v_{0}&\equiv&
    T^{(1,n,s,\epsilon)}\left.\frac{d^{2n+1}v}{dx^{2n+1}}
    \right|_{x_{0}}h^{2n}
    +O(h^{2n+1})\,,
    \\
    \left.\frac{d^{2}v}{d^{2}x}\right|_{x_{0}}-D^{(2,n)}
    &\equiv&T^{(2,n)}
    \left.\frac{d^{2n+2}v}{dx^{2n+2}}\right|_{x_{0}}h^{2n}+O(h^{2n+2})\,.
  \end{eqnarray*}
  A direct calculation yields
  \begin{eqnarray*}
  T^{(1,n,s,1)}=T^{(1,n,s,-1)} = T^{(1,n,s)}&=&(-1)^{s+n}\frac{(n+s)!(n-s)!}{(2n+1)!}\,,\\
    T^{(2,n)}&=&d_{n}=(-1)^n\frac{2(n!)^2}{(2n+2)!}\,.
  \end{eqnarray*}
  It is well known that the centered FDO has the smallest leading order 
  truncation error,
  $\left\vert T^{(1,n,0)}\right\vert<\left\vert T^{(1,n,s)}
  \right\vert$ for $s>0$ (see also table \ref{TableTruncErrorsD1}).
  
  \begin{table}
    \center
    \begin{tabular}{c|c|c|c|c}
      $\left\vert T^{(1,n,s)}\right\vert$
      & n=1 & n=2 & n=3 & n=4\tabularnewline
      \hline
      &&&\tabularnewline
      s=0 & $\frac{1}{6}$ & $\frac{1}{30}$ & $\frac{1}{140}$ 
      & $\frac{1}{630}$\tabularnewline
      &&&\tabularnewline
      s=1 & $\frac{1}{3}$ & $\frac{1}{20}$ & $\frac{1}{105}$
      & $\frac{1}{504}$\tabularnewline
    \end{tabular}
    \caption{Leading order truncation errors for the first order discrete
      derivative.}
    \label{TableTruncErrorsD1}
  \end{table}
  \subsection{Fourier representation of difference operators}
  \label{sectFourier}
  We assume a finite grid defined by a set of $N$ points, 
  \begin{equation}
    \mathcal{S}_{\underline{x}}(N)=\{x_{\nu }=\nu h,\,\text{ with } h=2\pi/N,\,
    \forall \nu =0,\dots,N-1\}\,,
   \label{defSgridpoints1D}
  \end{equation}
  and consider periodic grid functions $v_{\nu }=v_{\text{mod}(\nu ,N)}$, 
  decomposed as
  \begin{equation}
    v_{\nu }=\sum_{\omega\in \mathcal{S}_{\underline{\omega}}(N)}
    \hat{v}(\omega)b_{\nu}(\omega)\,,
    \label{FourierRepresGridFunc1D}
   \end{equation}
  where 
  \begin{eqnarray}
    b_{\nu}(\omega)&=&(2\pi)^{-1/2}e^{i\omega x_{\nu}}\,,\\
   \mathcal{S}_{\underline{\omega}}(N) &=&\left\{
    \begin{array}{cc}
      \{-N/2+1,\dots,N/2\},& \text{if}\quad\text{N is even}\\
      \{-(N-1)/2,\dots,(N-1)/2\},& \text{if}\quad\text{N is odd}\,.
    \end{array}
    \right.
    \label{defSomega1D}
  \end{eqnarray}
  The set $\mathcal{S}_{\underline{\omega}}(N)$ 
  represents the set of discrete wave numbers, and in 
  the space of periodic grid functions   
  the set $\{b_{\nu}(\omega),\,\omega\in\mathcal{S}_{\underline{\omega}}(N)\}$ 
  forms a orthonormal basis with respect to the scalar product 
  and the associated norm
  \begin{equation}
    (v,u)_{h}=
    \sum_{x_{\nu }\in\mathcal{S}_{\underline{x}}(N)}v^{\dagger}(x_{\nu }) u(x_{\nu })V_{h},
    \quad
    \left\Vert v \right\Vert_{h}^{2}=(v,v)_{h}.
    \label{normh1d}
  \end{equation}
  with $V_{h}=h$.
  The quantities $\hat{v}(\omega)$ represent the \textit{discrete 
    Fourier coefficients}.
  The scalar product satisfies the Parseval relation:
  \begin{equation}
    (v,u)_{h}=\sum_{\omega\in \mathcal{S}_{\underline{\omega}}(N)}
    \hat{v}^{\dagger}(\omega)
    \hat{u}(\omega).
    \label{Parseval1d}\end{equation}
    Let $\xi=\omega h\in \mathcal{S}_{\underline{\xi}}(N)$ with
    \begin{equation}
      \mathcal{S}_{\underline{\xi}}(N)=\left\{
      \begin{array}{cc}
        \{-\pi+2\pi/N,\dots,\pi\},& \text{if}\quad\text{N is even}\\
        \{-\pi+\pi/N,\dots,\pi-\pi/N\},& \text{if}\quad\text{N is odd}
     \end{array}\right .\,.
      \label{defSxi1D}
    \end{equation}

    Now apply the shift operator $S^{k}$ on a basis vector $b_{\nu }(\omega)$. 
    This leads to 
    \begin{equation}
    S^{k}e^{i\omega x_{\nu }}=\hat{S}^{k}(\xi)e^{i\omega x_{\nu }}, 
    \quad \text{with}\quad
    \hat{S}^{k}(\xi)=e^{i\xi k}\,.\nonumber
  \end{equation}
  The function $\hat{S}^{k}(\xi)$ represents the \textit{discrete 
    Fourier symbol} of the shift operator.
  For any discrete operator $D=\sum_{k}a^{k}(h)S^{k}$ the 
  Fourier symbol is defined by
  \begin{equation}
    D e^{i\omega x_{\nu }}=\hat{D}(\xi;h)e^{i\omega x_{\nu }}, \quad \text{with}\quad
    \hat{D}(\xi;h)=\sum_{k}a^{k}(h)\hat{S}^{k}(\xi)\,,\nonumber
  \end{equation}
  and for a general finite difference operator $D^{(m,n,s,\epsilon)}$ 
  the symbol is
  \begin{equation}
    \hat{D}^{(m,n,s,\epsilon)}(\xi;h)=h^{-m}\sum_{k=-n+\epsilon s}^{n+\epsilon s}
    \tilde{f}_{m,n,s,\epsilon,k}e^{i\xi k}\,.
  \end{equation}
  For the elementary discrete operators (\ref{elemoperators}) we obtain
    \begin{eqnarray*}
      \hat{D}_{\pm}(\xi;h)&=&\pm h^{-1}(e^{\pm i\xi}-1), \\
      \hat{\delta}_{0}(\xi)&=&i\hat{\delta}(\xi),
      \qquad\text{ where } \hat{\delta}(\xi)\equiv\sin \xi,\\
      \hat{p}(\xi)&=&-\hat{\Omega}^{2}(\xi),
      \quad\text{where } 
      \hat{\Omega}(\xi)\equiv2\sin\frac{\xi}{2}\,,
    \end{eqnarray*}
    and it is useful to note that 
    $
      \left\vert \hat{D}_{+}(\xi;h) \right\vert=
      \left\vert \hat{D}_{-}(\xi;h) \right\vert
      =h^{-1}\hat{\Omega}(\xi)
   $.

      The symbols for the first and second order derivative operators
      are straightforwardly computed using (\ref{FDOconvenient}),
    \begin{eqnarray}
      \hat{D}^{(1,n)}(\xi;h)&=&i\hat{d}^{(1,n)}(\xi)h^{-1}\,,
      \nonumber\\
      \hat{D}^{(2,n)}(\xi;h) &=&-\hat{d}^{(2,n)}(\xi)h^{-2}\,,
      \nonumber\\
      \hat{R}^{(n)}(\xi;h) &=&\hat{r}^{(n)}(\xi)h^{-2}
      \nonumber\\
      \hat{D}^{(1,n,s,\epsilon)}(\xi;h)
      &=&\left(\epsilon\,\hat{\mathbf{d}}^{(1,n,s)}(\xi)
      +i\hat{d}^{(1,n,s)}(\xi)\right)h^{-1}\,,
    \end{eqnarray}
    where we define
    \begin{eqnarray}
      \hat{d}^{(1,n)}&\equiv&\hat{\delta}
      \sum_{k=0}^{n-1}\left|c_{k}\right|\hat{\Omega}^{2k}\,,
      \nonumber\\
      \hat{d}^{(2,n)}&\equiv&
      \hat{\Omega}^{2}
      \sum_{k=0}^{n-1}\left|d_{k}\right|\hat{\Omega}^{2k}\,,
      \nonumber\\
      \hat{r}^{(n)}&=&-(\hat{d}^{(1,n)})^2+\hat{d}^{(2,n)}
      =\frac{n \left\vert c_{n-1}\right\vert}{2}\hat{\Omega}^{2(n+1)}
      \sum_{k=0}^{n-1}\frac{\left\vert c_{k}\right\vert}
          {n+1+k}\hat{\Omega}^{2k}\,,
          \nonumber \\
      \hat{\mathbf{d}}^{(1,n,s)}
      &=&	
      \hat{\Omega}^{2n+2}\sum_{k=0}^{s-1}
      (-1)^{n+k}b_{k}\hat{\Omega}^{2k}  \,,
      \nonumber\\
      \hat{d}^{(1,n,s)}&=&\hat{d}^{(1,n)}
      +\hat{\delta}\hat{\Omega}^{2n}
      \sum_{k=0}^{s-1}(-1)^{n+k}a_{k}
      \hat{\Omega}^{2k}
      \,. 
      \label{defFourierd1d2r}
    \end{eqnarray}
    In the following we list a series of particularly relevant
    properties of the Fourier symbols, further properties are given in appendix \ref{appPropFourier}.
    
 First note that the quantities $\hat{r}^{n}$ and $\hat{d}^{(2,n)}$ are 
 positive, and even more, from (\ref{defFourierd1d2r}) 
 it is easy to check that the following inequalities hold:
 \begin{eqnarray}
     &&0\leq\hat{d}^{(2,n)}\leq\hat{d}^{(2,n+1)}\,,
     \label{ineqd2}
     \\
     &&1\geq\frac{\hat{r}^{(n)}}{\hat{d}^{(2,n)}}\geq
     \frac{\hat{r}^{(n+1)}}{\hat{d}^{(2,n+1)}}\,.
     \label{ineqr}\\
   &&C_{n}^{-1}\hat{\Omega}^{2}
   \leq\hat{\Omega}^{2}\leq \hat{d}^{(2,n)}
   \leq C_{n}\hat{\Omega}^{2}, \quad\text{where}\quad 
   C_{n}\equiv 1+\sum_{k=1}^{n-1}\left|d_{k}\right|4^{k}\geq 1\,.
   \label{boundd2}
 \end{eqnarray}
 The real part of the Fourier symbol of the first derivative is
 an even function of the frequency $\xi$, while the imaginary part  
 is an odd function.
 The real part of the Fourier symbol of the first derivative also
 \begin{itemize}      
 \item vanishes for centered operators ($s=0$),
 \item keeps the same sign for all frequencies,
   in case the operator is one-point off-centered ($s=1$),
 \item changes sign for off-centering by more than one point ($s>1$).
 \end{itemize}
 The derivatives with respect to $\xi$ of the Fourier functions satisfy: 
 \begin{eqnarray}
    \partial_{\xi}\hat{d}^{(2,n)}
    =2\hat{d}^{(1,n)}\,,
   && \partial_{\xi}\hat{r}^{(n)}
    =2\frac{(n!)^{2}}{(2n)!}
   \hat{\Omega}^{2n}\hat{d}^{(1,n)}\,  \label{derivsymbols}\\
    \partial_{\xi}\hat{\mathbf{d}}^{(1,n,s)}
   =
   \frac{(-1)^s}{C_{2n}^{n-s}}
   \sin (s\,\xi)\hat{\Omega}^{2n}\,,
    &&
   \partial_{\xi}\hat{d}^{(1,n,s)}
   =1
   -\frac{(-1)^s}{C_{2n}^{n-s}}\cos (s\,\xi)
   \hat{\Omega}^{2n}\,.
 \nonumber
\end{eqnarray}
 In the following we show some plots to illustrate how the errors 
 of the Fourier symbols scale with the order of 
 approximation and  off-centering. The error is defined in respect 
 to the continuum limit, i.e., $h^{-m}(i\xi)^{m}$ for  
 $\hat{D}^{(m,n,s,\epsilon)}$.
 
\begin{figure}[pth]
   \centering
   \includegraphics[width=0.325\textwidth,height=0.18\textheight]
		   {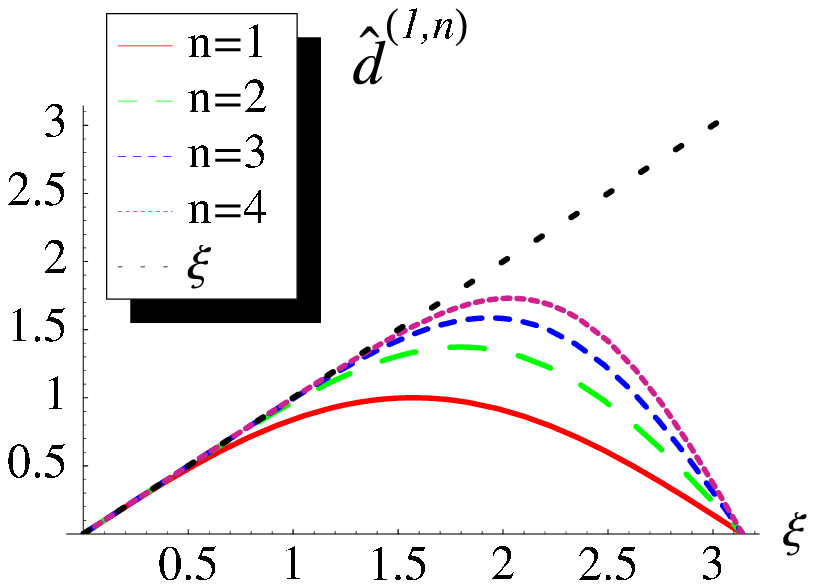}
                   \includegraphics[width=0.325\textwidth,height=0.18\textheight]
		                   {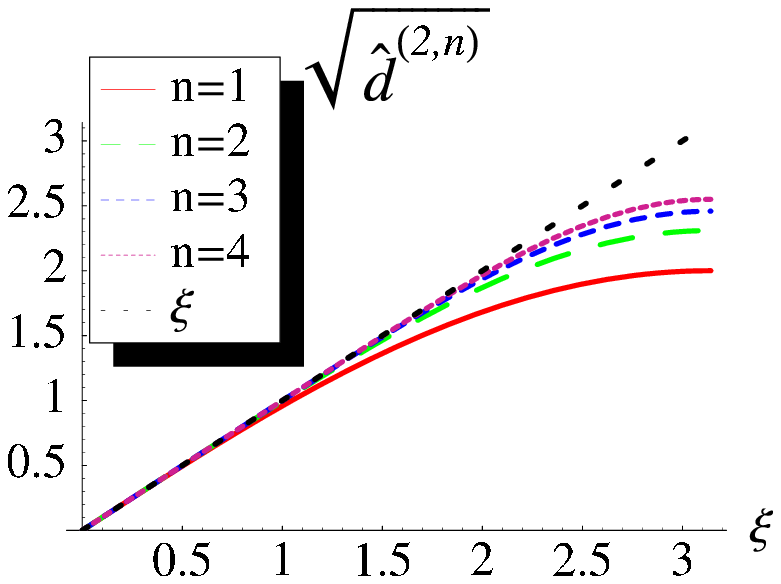}
                                   \includegraphics[width=0.325\textwidth,height=0.18\textheight]
		  {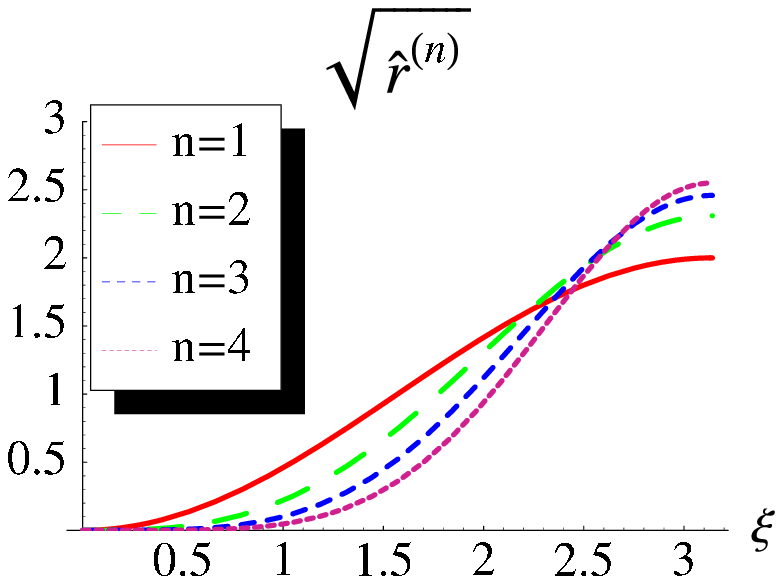}
		  \caption{
		    Fourier symbols 
                    as functions of the frequency $\xi$, 
		    for different approximation orders. For increasing
		    order the second derivative
		    becomes more accurate for all frequencies,
		    while the first derivative does not converge for
		    $\xi = \pi$.}
\label{plotD1D2}
\end{figure}

 Figure~\ref{plotD1D2} shows the Fourier symbols 
 $\hat{d}^{(1,n)},\hat{d}^{(2,n)}$
 and $\hat{r}^{(n)}$ as functions of the 
 frequency $\xi$ for different orders  of accuracy. For increasing
 approximation order, the second derivative
 becomes more accurate for all frequencies, 
 while the first derivative does not converge 
 to the continuum limit for the highest frequency in the grid, 
 where the symbol is zero. 
 The $\pi$-frequency will not be captured also by 
 the off-centered discrete operators associated with 
 the first derivative. In addition, for them, 
 the error scales with the order only at small frequencies.

\begin{figure}[ht]
  \centering
  \includegraphics[width=0.49\textwidth,height=0.19\textheight]{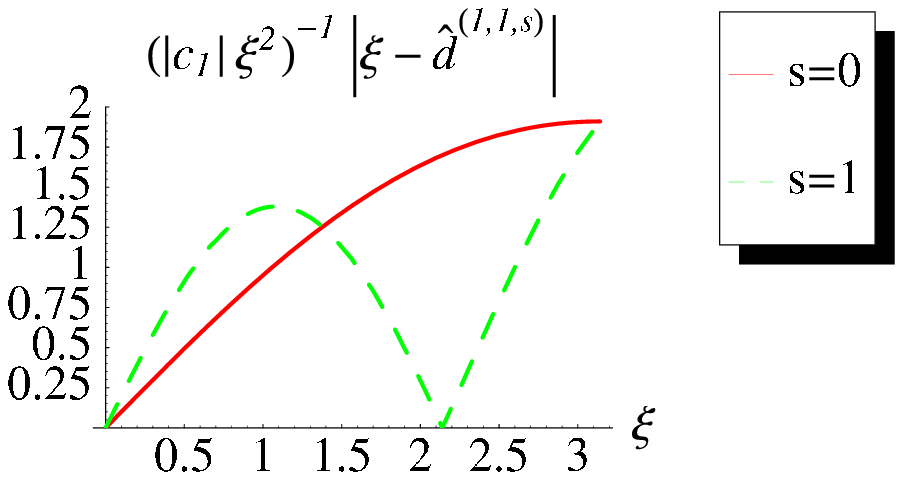}
  \includegraphics[width=0.49\textwidth,height=0.19\textheight]{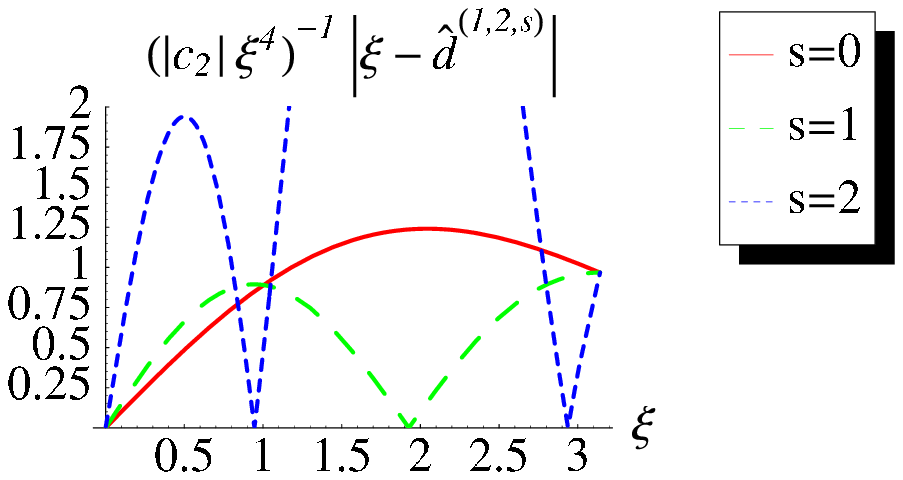}
  \includegraphics[width=0.49\textwidth,height=0.19\textheight]{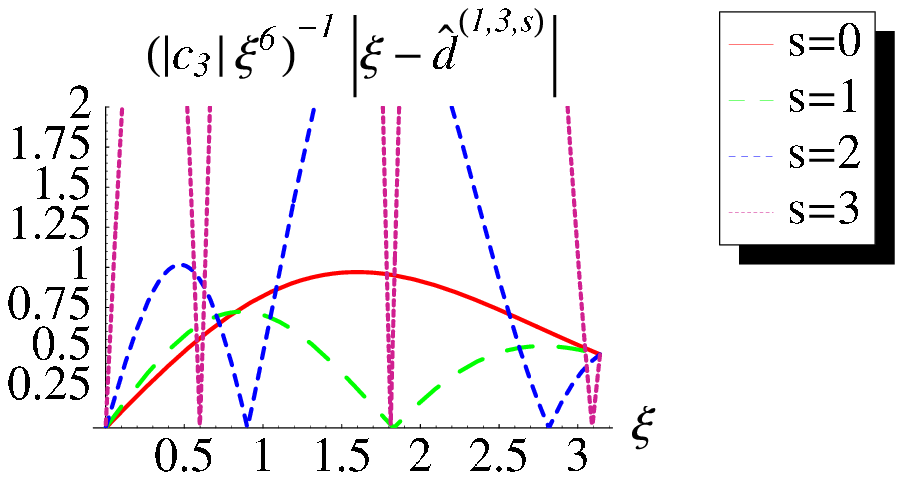}
  \includegraphics[width=0.49\textwidth,height=0.19\textheight]{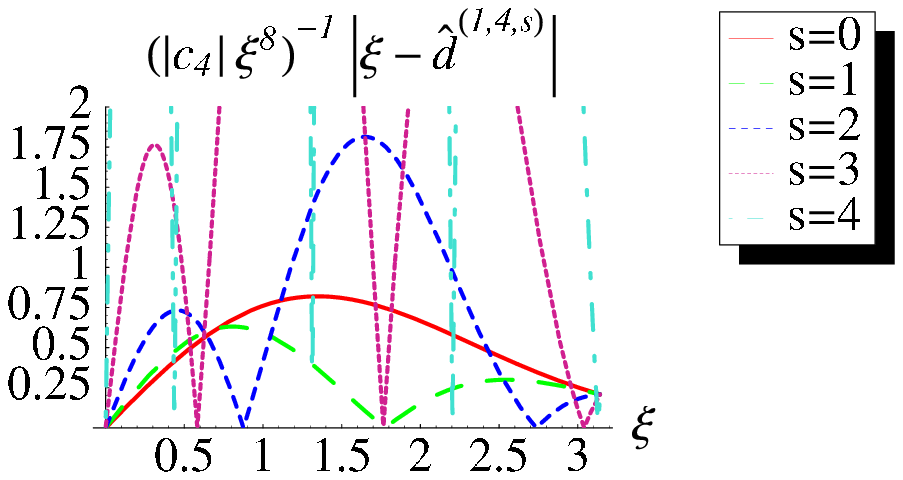}
  \caption{\textbf{Errors for $\hat{d}^{(1,n,s)}$ at 
      fixed order but different off-centerings}
    are shown, scaled with 
    $\left\vert c_{n}\right\vert\xi^{2n}$
    ($n=1,2,3,4$).
    For small $\xi$ the curves are straight lines
    with the slope $\left\vert c_{n}T^{(1,n,s)}\right\vert$, 
    and larger $s$ increases the error.
    At higher frequencies this behavior changes, but 
    for $s\geq 2$ large intervals appear
    where the error overcomes by far the error when $s=0$. 
    For $s=1$ we observe that while at small frequencies, 
    the error is slightly larger than for $s=0$, for each order
    there is a frequency $\xi^{(n)}$ beyond which the error is
    smaller than for $s=0$.
   }\label{plotscalederrorD1ns}
\end{figure}

 Figure~\ref{plotscalederrorD1ns} shows 
 the scaling of the error
 for $\hat{d}^{(1,n,s)}$ with the off-centering
 at fixed order of approximation. 
 In the region of small frequencies, off-centering increases the error.
 At larger frequencies this behavior changes.
 For each $s$, there are exactly $s$ frequencies in $(0,\pi)$ where 
 the error cancels. However, for $s\geq 2$ there are large intervals 
 where the error overcomes by far the error when $s=0$. 
 For $s=1$ we observe that while at small frequencies, 
 the error is slightly larger than for $s=0$, for each order $2n$,
 there is a frequency, $\xi^{(n)}$, beyond which the error is
 smaller than for the case $s=0$.
 This frequency can be computed numerically, e.g.~$\xi^{(1)}=1.3787$, 
 $\xi^{(2)}=1.0036$, $\xi^{(3)}=0.8234$, $\xi^{(4)}=0.7136$.
 
\subsection{Generalization to $d$-dimensions}
\label{subsec:GenTo_dDim}
In this work we will use the common straightforward generalization of 
finite difference operators from one to $d$ dimensions. We extend
first derivatives in a particular coordinate direction in the trivial way,
and second derivative operators in the $jl$-directions are 
defined as 
  \begin{equation}
    D^{(2,n)}_{jl}=\left\{
    \begin{array}{cc}
      D^{(1,n)}_{j}D^{(1,n)}_{l},& \quad\text{for}\quad j\neq l\\
      D^{(2,n)}_{j},& \quad\text{for}\quad j=l\,.
    \end{array}
    \right.
  \end{equation}
  The Fourier symbols for the first and second derivative 
  operators take the form
  \begin{eqnarray}
    \hat{D}^{(1,n,s,\epsilon)}_{j}&=&h_{j}^{-1}
    \left(\epsilon_{j}\,\hat{\mathbf{d}}^{(1,n,s_{j})}(\xi_{j})
    +i\hat{d}^{(1,n,s_{j})}(\xi_{j})\right)\,,
    \nonumber
    \\
   \hat{D}^{(1,n)}_{j}&=&h_{j}^{-1}i\hat{d}^{(1,n)}_{j}\,,
   \nonumber
   \\
   \hat{D}^{(2,n)}_{jl}&=&h_{j}^{-1}h_{l}^{-1}
   \left\{
   \begin{array}{cc}
     -\hat{d}^{(1,n)}_{j}\hat{d}^{(1,n)}_{l}&j\neq l\\\\
     -\hat{d}^{(2,n)}_{j}&j=l
   \end{array}
   \right.\,,
   \label{D1D2FourierSymbols_in_jDir}
 \end{eqnarray}
  In order to simplify notation, 
  we will also  use the following convention: 
  any function in frequencies, and possible, grid spacings,  
  $\hat{f}(\xi_{i_{1}},\dots \xi_{i_{r}};h_{i_{1}},\dots h_{i_{r}})$
  will be referred to by $\hat{f}_{i_{1}\dots i_{r}}$, 
  and by $\hat{f}$ in case it depends on all frequencies.
  More detailed definitions for the $d$-dimensional case are  given
  in appendix \ref{sec:GenTo_dDim}.
   
\subsection{Dissipation Operators}
In order to achieve numerical stability for problems that go beyond the
linear constant coefficient case, it is common practice to add 
artificial dissipation to the right-hand-sides of the time evolution equations.
In this work we only deal with the constant coefficient
problem, but in \ref{RoleDissip} we will also use dissipation to stabilize 
numerical schemes which would be unstable otherwise. 

Dissipation terms are typically chosen to converge away fast
enough so as not to change the convergence order of the scheme.
Here we use the Kreiss-Oliger dissipation operator $\mathcal{D}^{(2m)}$ 
of order $2m$ \cite{Kreiss} and its Fourier representation
 $\hat{\mathcal{D}}^{(2m)}$,
  \begin{equation}
   \mathcal{D}^{(2m)} =-\frac{(-1)^{m}}{2^{2m}}
    \sum_{j=1}^{d}\sigma_{j}h_{j}^{2 m-1}(D_{+j})^{m} (D_{-j})^{m},
\quad
   \hat{\mathcal{D}}^{(2m)}
    = -\frac{1}{2^{2m}}\sum_{j=1}^{d}\frac{\sigma_{j}}{h_{j}}
    \hat{\Omega}^{2m}_{j},
    \label{KOdiss}
  \end{equation}
  for a $2m -2$ accurate scheme, where the parameters 
  $\sigma_{j}\geq 0$ regulate the
  strength of the dissipation. Using this form of numerical 
  dissipation, theorems can be proved concerning the 
  numerical stability of non-constant-coefficient
  hyperbolic PDEs \cite{Kreiss}.
  Note that it is more common to have the dissipation 
  parameters $\sigma_j$ not depend
  on the direction or other parameters of the system.
  \section{Numerical stability for first order in time,
    second order in space hyperbolic systems}\label{sec:StabMethod}
  We now turn to the analysis of numerical stability for the 
  system (\ref{contsystem}), following \cite{CalHinHusa}. 
  This  problem is greatly
  simplified by adopting the method-of-lines approach
  where initially time is kept continuous and only space is discretized 
  (i.e. the semi-discrete problem).
  Then the discrete system to be analyzed becomes  
  \begin{equation}
    \frac{d}{dt}v=P v,\,\,\,
    v=(U,V)^{T} \,,
  \nonumber\end{equation}
  
  \begin{eqnarray}  
    P&=&\left(
    \begin{array}{cc}
      A^{j}D_{j}^{(1,n)}+B&C\\
      D^{jl}D_{jl}^{(2,n)}+E^{j}D_{j}^{(1,n)}+F&G^{j}D_{j}^{(1,n)}+J
    \end{array}
    \right)\,.
    \label{DiscreteSyst}
  \end{eqnarray}
  We consider periodic grid functions in each direction, 
  and Fourier transform the system as discussed in 
  appendix \ref{sec:GenTo_dDim}.
  Then a first order reduction is performed 
  by introducing the variable $\hat{w}$,
  \begin{equation}
    \hat{w}\equiv i \Omega_{0}\hat{u},\,\,\Omega_{0}^{2}=\sum_{j=1}^{d}
    \left|\hat{D}_{+j}\right|^{2} \,,
    \label{DefW}
  \end{equation}
  where $\hat{D}_{+j}$ is the Fourier symbol of the 
  forward finite difference operator in the $j$-direction, $D_{+j}$.
  The case $\Omega_{0}= 0$ (which corresponds to zero 
  frequencies in all directions) does not play any role 
  in the stability analysis\footnote{the zero frequency vector 
    corresponds to a term constant in space.}, so we  define 
  $\mathcal{S}_{\underline{\xi}}^{*}(N)
  =\mathcal{S}_{\underline{\xi}}(N)- {\mathbf{0}^{d}}$
  and assume $\xi\in \mathcal{S}_{\underline{\xi}}^{*}(N)$.

  By (\ref{DefW}) we obtain the following system of ODEs 
  \begin{eqnarray}
    \frac{d}{dt}\hat{v}_{R}
    &=&\hat{P}_{R}\hat{v}_{R} \text{ with }
    \,\,\,\hat{v}_{R}=(\hat{u},\hat{w},\hat{v})^{T}\,,\nonumber\\
    \hat{P}_{R}&=&\left(
    \begin{array}{ccc}
   B& (i\Omega_{0})^{-1}A^{j}\hat{D}_{j}^{(1,n)}&C\\
   0& A^{j}\hat{D}_{j}^{(1,n)}+
   B&i\Omega_{0} C\\
   F&(i\Omega_{0})^{-1}\left(D^{jl}\hat{D}_{jl}^{(2,n)}+E^{j}\hat{D}_{j}^{(1,n)}\right)
   &G^{j}\hat{D}_{j}^{(1,n)}+J
    \end{array}
    \right)\,.
  \end{eqnarray}
  Using the theorem 5.1.2 of \cite{Kreiss} CHH show that 
  the terms which correspond 
  to the continuum lower order terms
  can be dropped from $\hat{P}_{R}$ without affecting the stability
  analysis if   
  \begin{equation}
    (i\Omega_{0})^{-1}\hat{D}_{j}^{(1,n)}, 
    \quad
    k\hat{D}_{j}^{(1,n)},
    \quad
    k\Omega_{0}^{-1}\hat{D}_{jl}^{(2,n)}
    \label{symbols1}
  \end{equation}
  are bounded for all frequencies $\xi\in \mathcal{S}_{\underline{\xi}}^{*}(N)$.
  In the relations (\ref{symbols1}),  
  $k$ represents the time step.
  We  will show in lemma \ref{boundlot} that this is indeed 
  the case for any order of accuracy $2n$. 
 
  Having proved this, the rest of the discussion in CHH applies.
  The problem now reduces to the analysis of a 
  first order system with the principal part:
  \begin{eqnarray}
  \hat{P'}_{R}&=&\left(
    \begin{array}{cc}
       A^{j}\hat{D}_{j}^{(1,n)}&i\Omega_{0} C\\
      (i\Omega_{0})^{-1}D^{jl}\hat{D}_{jl}^{(2,n)}
       &G^{j}\hat{D}_{j}^{(1,n)}
    \end{array}
    \right)\,.
  \end{eqnarray}
  For this type of system, sufficient conditions 
  for stability have been deduced in \cite{Kreiss}. 
  These conditions have been 
  exploited in CHH to analyze the stability of 
  the second order system.
  By introducing the so-called 
  \textit{second-order principal symbol} of the semi-discrete system,
  \begin{equation}
    \hat{P}'=\left(
    \begin{array}{cc}
      A^{j}\hat{D}_{j}^{(1,n)}&C\\
      D^{jl}\hat{D}_{jl}^{(2,n)} &G^{j}\hat{D}_{j}^{(1,n)}
    \end{array}
    \right)\,,
  \end{equation}
  and assuming that the time integration 
  is done using one-step explicit schemes,  
  CHH show that the following conditions are 
  sufficient for stability:
  \vspace{0.5cm}
  
  \textbf{Condition 1:} There exists a hermitian matrix 
  $\hat{H}(\xi,h)$ such that 
  \begin{eqnarray}
      &&K^{-1}I_{\Omega_{0}}\leq\hat{H}\leq KI_{\Omega_{0}},\,\, 
      I_{\Omega_{0}}=\text{diag}[\Omega_{0}^{2},I_{qN}],\nonumber\\
      &&\hat{H}\hat{P}'+\hat{P}'^{\dagger}\hat{H}=0,
      \label{Symmetrizer}
  \end{eqnarray}
  for some positive constant $K$.
  
  \textbf{Condition 2:} The eigenvalues of $k\hat{P}'$ 
  have non-positive real parts and 
  \begin{equation}
    \sigma(k\hat{P}')\leq\alpha_{0}\,
    \label{LocStabCond}
  \end{equation}
  where $\sigma(k\hat{P}')$ is the maximum 
  spectral radius of $k\hat{P'}$  and 
  $\alpha_{0}$ is a constant specific to the time integrator.

  \textbf{Remarks:}    
  \begin{itemize}
  \item  The condition (\ref{Symmetrizer}) implies that the
    semi-discrete problem is stable with respect to
    the norms $D_{\pm}$ defined as:
    \begin{equation}
      \left\Vert v\right\Vert^{2}_{h,D_{\pm}}=
      \sum_{i=1}^{d}
      \left\Vert D_{\pm i}U\right\Vert^{2}_{h}
      +\left\Vert V\right\Vert^{2}_{h}\,.
      \end{equation}
    where $\left\Vert . \right\Vert^{2}_{h}$ is the $d$-dimensional 
    analog of (\ref{normh1d}).
  \item  The semi-discrete problem is stable also in the norm 
    $\left\Vert v\right\Vert_{h,H}^{2}$  defined by:
    \begin{equation}
      \left\Vert v\right\Vert_{h,H}^{2}= 
      \sum_{\omega \in \mathcal{S}_{\underline{\omega}}(N)}\hat{v}^{\dagger}
      \hat{H}\hat{v}\,.
      \label{normHh}
    \end{equation}
    This norm is conserved by the principal symbol of the evolution system, 
    $\left\Vert v(t,.)\right\Vert_{h,H}= \left\Vert v(0,.)\right\Vert_{h,H}$. 

  \item The constant $\alpha_{0}$ in (\ref{LocStabCond}) 
    denotes  the radius of local stability on the imaginary axis 
    ($R_{lsia}$)
    in case the eigenvalues of $k\hat{P}'$ are purely imaginary, 
    and  the radius of local stability ($R_{ls}$), otherwise.
    \footnote{for the classical fourth order Runge-Kutta, 
      $R_{lsia}=\sqrt{8}=2.83$ and $R_{ls}=2.61$.
    } 
  \item In case all the grid spacings are equal, 
    $h_{1}=\dots= h_{d}=h$, 
    and we introduce the Courant factor $\lambda=k/h$, 
    then the relation (\ref{LocStabCond}) 
    provides the Courant limit: 
    \begin{equation}
      \lambda\leq\frac{\alpha_{0}}{\sigma(h\hat{P}')}\,.
      \label{CourantLimit}
    \end{equation}
  \item If the right hand side of the system (\ref{DiscreteSyst}) is  
    modified  by adding artificial dissipation (using the operator 
    $\mathcal{D}^{(2m)}$ defined in (\ref{KOdiss}))
    and/or by adding shift advection terms of the form 
    $I\beta^{j}{D}_{j}^{(1,n,s_{j},\epsilon)}$ 
    (where ${D}_{j}^{(1,n,s_{j},\epsilon)}$ is the non-centered 
    FDO in the $j$-direction constructed from (\ref{FDOconvenient})), 
    these modifications only have effect
    on the diagonal entries of the principal part.
    The new system will have different eigenvalues than $\hat{P}'$
    but the same set of eigenvectors.
    The symmetrizer will not depend on
    the way we discretize the advection terms, 
    nor on the dissipation operator.   
    The stability \textbf{Conditions 1-2} remain valid if 
    \begin{equation}
      (i\Omega_{0})^{-1}\hat{D}_{j}^{(1,n,s_{j},\epsilon)},\quad 
      (i\Omega_{0})^{-1}\hat{\mathcal{D}}^{(2m)}
      \label{symbols2}
    \end{equation}
    are bounded and this will be shown below together 
    with the boundedness of the terms (\ref{symbols1}).
  \end{itemize}

\begin{lemma}\label{boundlot}
  The following quantities are bounded for all 
  frequencies $\xi\in \mathcal{S}_{\underline{\xi}}^{*}(N)$.
  \begin{equation}
    (i\Omega_{0})^{-1}\hat{D}_{j}^{(1,n)},~~
    k\hat{D}_{j}^{(1,n)},~~
    k\Omega_{0}^{-1}\hat{D}_{jl}^{(2,n)},~~
    (i\Omega_{0})^{-1}\hat{D}_{j}^{(1,n,s_{j},\epsilon)},~~
    (i\Omega_{0})^{-1}\hat{\mathcal{D}}^{(2m)}
    \label{symbols}
  \end{equation}
\end{lemma}
Making use of the relations (\ref{D1D2FourierSymbols_in_jDir})
and (\ref{KOdiss}), the proof reduces 
to showing the boundedness of
$$\hat{\Omega}_{0}^{-1}\hat{d}_{j}^{(1,n)},\quad 
\hat{d}_{j}^{(1,n)},\quad
\hat{\Omega}_{0}^{-1}\hat{d}_{j}^{(2,n)},\quad
\hat{\Omega}_{0}^{-1}(\hat{d}_{j}^{(1,n)})^{2},$$
$$\hat{\Omega}_{0}^{-1}\hat{d}^{(1,n,s_{j})}_{j},\quad
\hat{\Omega}_{0}^{-1}\hat{\mathbf{d}}^{(1,n,s_{j})}_{j},\quad
\hat{\Omega}_{0}^{-1}\hat{\Omega}^{2m}_{j}\,.
$$
From the relations (\ref{defFourierd1d2r}) 
we observe that each of these quantities 
can be written formally as a product 
$\hat{\Omega}_{0}^{-1}\hat{\Omega}_{j}F(\xi_{j})$,
with $F(\xi_{j})$ a continuous and bounded
function in $(-\pi,\pi]$. 
  Since $\hat{\Omega}_{0}^{-1}\hat{\Omega}_{j}$ 
  is bounded for all $\Omega_{j}\in(-2,2]$, $j=1,\dots, d$
  but not all zero in the 
  same time, we obtain the desired result.
\section{Application: Scalar Wave Equation}
\label{sectionWave}
\subsection{Semi-discrete Problem}
The system (\ref{contwaveeq3D}) is discretized assuming,
for simplicity, that the grid spacings are equal
($h_{1}=\dots=h_{d}=h$). The case $h_{i}\neq h_{j}$
for some directions $i$ and $j$ does not introduce
further complications in the following analysis.

We construct the semi-discrete system corresponding to 
(\ref{contwaveeq3D}) by: 
  \begin{eqnarray}
    \frac{d}{dt} \Phi &=&\sum_{j=1}^{d}
    \beta^{j}D^{(1,n,s_{j},\epsilon_{j})}_{j}\Phi+K
    \,, 
    \nonumber\\ 
    \frac{d}{dt}K&=&\gamma^{jl}D^{(2,n)}_{jl}\Phi
    +\sum_{j=1}^{d}\beta^{j}D^{(1,n,s_{j},\epsilon_{j})}_{j}K
    \,.\label{SemiDisWaveEq3D}
  \end{eqnarray}
  This way of discretizing the first order derivative terms, which correspond
  to advection along the shift vector $\beta^i$, with off-centered derivatives
  has become customary in numerical relativity (see e.g.
  \cite{Alcubierre:2000yz,Zlochower:2005bj,Husa:2007hp}).
  
  We define the shorthand quantity $\hat{\Delta}$ as
  $$
   \hat{\Delta}\equiv \sqrt{-\gamma^{jl}\hat{D}^{(2,n)}_{jl}}
   =h^{-1}
   \sqrt{\gamma^{jl}\hat{d}^{(1,n)}_{j}\hat{d}^{(1,n)}_{j}
     +\sum_{j=1}^{d}\gamma^{jj}\hat{r}^{(n)}_{j}}\,.
   $$
  Then the discrete symbol, the diagonalizing matrix and the eigenvalues 
  can be written as
\begin{equation}
  \hat{P}'=\left(
  \begin{array}{cc}
    \beta^{j}\hat{D}^{(1,n,s,\epsilon)}_{j}
    &1\\
    -\hat{\Delta}^{2}&
    \beta^{j}\hat{D}^{(1,n,s,\epsilon)}_{j}
  \end{array}
  \right)\,,
\end{equation}
\begin{equation}
  \begin{array}{cc}
    \hat{T}^{-1}=\left(
    \begin{array}{cc}
      i\hat{\Delta}&1\\
      -i\hat{\Delta}&1
    \end{array}
    \right),\quad
    &
    \hat{\Lambda}_{\pm}=
    \sum_{j=1}^{d}\beta^{j}\hat{D}^{(1,n,s_{j},\epsilon_{j})}_{j}
    \pm i\hat{\Delta}\,.
  \end{array}
\label{evandrest}
\end{equation}  
Because $\hat{r}^{(n)}_{j}\geq 0$, according to (\ref{defFourierd1d2r}), and 
the  matrix $\gamma^{jl}$ is positive definite, 
the quantity $\hat{\Delta}$ is real and
$\hat{\Delta}\geq 0$ with equality only when all $\xi_{j}$ are zero.
Thus 
\begin{equation}
  \hat{H}\equiv
  \frac{1}{2}\hat{T}^{-1\dagger}\hat{T}^{-1}=\left(
  \begin{array}{cc}
    \hat{\Delta}^{2}&0\\
      0&1
  \end{array}
  \right)
\nonumber\end{equation}
is a symmetrizer for the system (\ref{SemiDisWaveEq3D}).
We observe that the symmetrizer does not depend on the diagonal entries
of the symbol $\hat{P}'$, e.g. does not depend on the way we advect
the shift terms.

We still have to prove that there exists a constant $K\geq1$ such that
\begin{equation}
  K^{-1}\Omega_{0}^{2}\leq\hat{\Delta}^{2}\leq K\Omega_{0}^{2}\,.
  \label{reldelta}
\end{equation}
The positivity of the matrix $\gamma^{jl}$ implies 
the existence of a constant $c_{1}>0$ such that
\begin{equation}
  c_{1}\leq \min\gamma^{jj}\quad\text{and}\quad
  \gamma^{jl}y_{j}y_{l}\geq c_{1}\left\vert y\right\vert^{2}, 
\quad \forall y=(y_{1},\dots, y_{d})\in\Real^{d}\,.
  \label{relc1}
\end{equation}
Furthermore, because 
$\left|\gamma^{jl}\right|<\infty$ there also exists a 
constant $c_{2}>0$ such that
\begin{equation}
 c_{2}\geq \max\gamma^{jj}\quad\text{and}\quad  
\gamma^{jl}y_{j}y_{l}\leq c_{2}\left\vert y\right\vert^{2}, 
\quad \forall y=(y_{1},\dots, y_{d})\in\Real^{d}\,.
\label{relc2}
\end{equation}
Using (\ref{relc1}) and the inequalities (\ref{boundd2}) we obtain
\begin{equation}
  h^{2}\hat{\Delta}^{2}\geq
  (\min{\gamma^{jj}})\sum_{j=1}^{d}\hat{r}^{(n)}_{j}
  +c_{1}\sum_{j=1}^{d}(\hat{d}_{j}^{(1,n)})^{2}
  \geq c_{1}\sum_{j=1}^{d}\hat{d}^{(2,n)}_{j}
  \geq c_{1}\hat{\Omega}_{0}^{2}\,.
 \nonumber\end{equation}
 On the other hand, by (\ref{relc2}) and again (\ref{boundd2}) we have that
 \begin{equation}
   h^{2}\hat{\Delta}^{2}\leq (\max\gamma^{jj})\sum_{j=1}^{d}\hat{r}^{(n)}_{j}
   +c_{2}\sum_{j=1}^{d}(\hat{d}_{j}^{(1,n)})^{2}
   \leq c_{2}\sum_{j=1}^{d}\hat{d}^{(2,n)}_{j}
   \leq c_{2}C_{n}\hat{\Omega}_{0}^{2}\,.
   \nonumber\end{equation}
   We chose $K=\max\{c_{1}^{-1},(c_{2}C_{n}),1\}$ and obtain the relation 
   (\ref{reldelta}).\\

   The conserved discrete quantity in physical space associated 
   to  $\hat{H}$, i.e. the norm $\left\Vert v\right\Vert_{h,H}$
   defined in (\ref{normHh}), is 
   $$
    \left\Vert v\right\Vert_{h,H}^2
    =\frac{1}{h^{2}}\left[
      \sum_{j=1}^{d}
    \gamma^{jj}\sum_{k=1}^{n}\left|d_{k-1}\right|
    \left\Vert (hD_{+j})^{k}\Phi\right\Vert^{2}_{h}
    +
    \sum_{j\neq l}\gamma^{jl}
    \left\Vert hD_{l}^{(1,n)}\Phi\right\Vert^{2}_{h}
    \right]
    +\left\Vert K\right\Vert^{2}_{h},
  $$
where $v=(\Phi^{T},K^{T})^{T}$.
Having proved the existence of a symmetrizer we 
have proved that the semi-discrete problem is stable
with respect to the norms $D_{+}$ and $H$. 
Note again that the stability property does in particular not depend on
how the shift terms are discretized.

\subsection{Courant Limits and the Role of Dissipation}
\label{RoleDissip}
In order for the fully discrete problem to be stable we
impose the non-positivity condition on the real part of the eigenvalues 
and restrict the Courant factor $\lambda$ 
according to the inequality (\ref{CourantLimit}):
\begin{eqnarray}
  Re(\hat{\Lambda}_{\pm})&\leq& 0\,,
  \label{negev}
  \\ 
  \lambda&\leq&\frac{\alpha_{0}}
	 {\underset{\xi\in S_{\underline{\xi}}}{\max}
	 \left|h\hat{\Lambda}_{\pm}\right|}\,.
\end{eqnarray}
From (\ref{evandrest}) we have 
\begin{equation}
  h \RealPart (\hat{\Lambda}_{\pm})= \sum_{j=1}^{d}\beta^{j}
  \epsilon_{j}\hat{\mathbf{d}}^{(1,n,s_{j})}_{j}\,.
  \label{RealPartEV}
\end{equation}
The relation (\ref{negev}) has to hold for 
all frequencies $\xi_{j}\in(-\pi,\pi)$. 
Because each term $j$ in the sum (\ref{RealPartEV}) 
can be canceled individually at $\xi_{j}=0$, 
the non-positivity condition has to applied for each term.
The problem reduces to the study of the one-dimensional case, 
\begin{equation}
  \beta \epsilon \hat{\mathbf{d}}^{(1,n,s)}(\xi)\leq 0
  \quad\text{with} \quad\xi\in(-\pi,\pi]
    \label{RealPartEV_1d}
\end{equation}
Because $\hat{\mathbf{d}}^{(1,n,s)}$ is zero for 
$s=0$, negative for $s=1$ and changes sign for 
$s\geq 2$, it is clear that the condition holds 
for centered and one-point upwinded 
($\epsilon=\sign{\beta}$) schemes and is violated 
in all the other cases.

However, the condition can be reestablished 
if appropriate artificial dissipation is added to the system.
By using the Kreiss-Oliger dissipation operator (\ref{KOdiss}), 
the condition (\ref{RealPartEV_1d}) changes to 
\begin{equation}
  \beta \epsilon \hat{\mathbf{d}}^{(1,n,s)}(\xi)
  -\frac{1}{2^{2(n+1)}}\sigma \hat{\Omega}^{2(n+1)}(\xi)
  \leq 0
  \quad\text{with} \quad\xi\in(-\pi,\pi]
\end{equation}
This imposes a lower limit on the dissipation parameter $\sigma$: 
\begin{equation}
  \sigma\geq \sigma_{\min}(\beta,n,s,\epsilon)=
  \left\{
  \begin{array}{cc}
    2^{2(n+1)}\left\vert\beta\right\vert\bar{\sigma}_{+}^{(n,s)},
    &
    \epsilon=\sign{\beta}  \text{  (upwind )}  \\\\
    2^{2(n+1)}\left\vert\beta\right\vert\bar{\sigma}_{-}^{(n,s)},
    &
    \epsilon=-\sign{\beta} \text{  (downwind)}
  \end{array}
  \right.
  \label{minsigma}
\end{equation}
where we have denoted
\begin{equation}
  \bar{\sigma}_{+}^{(n,s)}\equiv
  \underset{\hat{\Omega} \in(0,2]}{\max}\frac{\hat{\mathbf{d}}^{(1,n,s)}}
    {\hat{\Omega}^{2(n+1)}}
    ,\quad
    \bar{\sigma}_{-}^{(n,s)}\equiv-
    \underset{\hat{\Omega}\in(0,2]}{\min}\frac{\hat{\mathbf{d}}^{(1,n,s)}}
      {\hat{\Omega}^{2(n+1)}}.
      \end{equation}
and used the fact that $\hat{\mathbf{d}}^{(1,n,s)}$ is,  
according to (\ref{defFourierd1d2r}), a sum over powers of $\hat{\Omega}$.

In table \ref{TableSigmas} we give the formulas 
for $\bar{\sigma}_{\pm}^{(n,s)}$ for $s=1,2,3$.
We remark that for all $n\geq 1$,  
$\bar{\sigma}_{+}^{(n,1)}<0$, $\bar{\sigma}_{-}^{(n,1)}>0$ 
and 
$\bar{\sigma}_{\pm}^{(n,s)}>0$ for all $s>1$.

This means that when using one-point upwinded stencils, 
we can add ``negative'' dissipation
and still obtain a stable scheme. In fact, the following
situations are equivalent:
\begin{itemize}
\item Upwind one point and add
  dissipation with $\sigma=
  2^{2(n+1)}\left\vert\beta\right\vert
  \bar{\sigma}_{+}^{(n,1)}<0$.
	\item Downwind one point and add
	  dissipation with $\sigma=
	  2^{2(n+1)}\left\vert\beta\right\vert
	  \bar{\sigma}_{-}^{(n,1)}>0$.
	\item Use the CFDO operator 
            $1/2\left(D^{(1,n,s,1)}+D^{(1,n,s,-1)}\right)$ 
          (constructed with $2(n+s)+1$ points), 
          and do not add dissipation, $\sigma=0$.
	\end{itemize}
      In any of the above three situations, the real part of the eigenvalues
      is zero, so if the Courant limit is small enough then we obtain
      stability on the imaginary axis.
      \begin{table}
	\center
      \begin{tabular}{c|c|c}
	& $\bar{\sigma}_{+}^{(n,s)}$ &$ \bar{\sigma}_{-}^{(n,s)}$
	\tabularnewline
	\hline
        &&\tabularnewline
	s=1&$-\frac{1}{2C_{2n}^{n+1}}\frac{1}{n+1} $&
	$\frac{1}{2C_{2n}^{n+1}}\frac{1}{n+1}$
	\tabularnewline
        &&\tabularnewline
	s=2&
	$\frac{1}{2C_{2n}^{n+2}}\frac{2}{n+1}$&
	$\frac{1}{2C_{2n}^{n+2}}
	\frac{2 n}{n^2+3 n+2}$
	\tabularnewline
        &&\tabularnewline
	s=3&
	$\frac{1}{2C_{2n}^{n+3}}\frac{n (n+4)}{(n+1) (n+2)^2}$&
	$\frac{1}{2C_{2n}^{n+3}}\frac{3}{n+1}$
	\tabularnewline
      \end{tabular}
      \caption{Formulas for the dissipation
	parameters $\bar{\sigma}_{\pm}^{(n,s)}$ when $s=1,2,3$.
	The quantity $2^{2(n+1)}\left\vert\beta^{j}\right\vert
	\bar{\sigma}_{\pm}^{(n,s)}$ ($\pm$ stands for upwind/downwind)
	represents the minimum dissipation that one has to add
	to make the numerical scheme stable.}
      \label{TableSigmas}
      \end{table}
      Now, coming back to the $d$-dimensional case, 
      it is obvious that the dissipation parameters 
      in (\ref{KOdiss}), $\sigma_{j}$, have to be chosen 
      according to  the value of the shift and the type of 
      off-centering in the $j$-direction 
      ($\sigma_{j}\geq  \sigma_{\min}(\beta_{j},n,s_{j},\epsilon_{j})$). 
      It can shown that, by choosing exactly 
      $\sigma_{j}= \sigma_{\min}(\beta_{j},n,s_{j},\epsilon_{j})$ 
      the Courant limit is maximized.
      However, to compute it explicitly, 
      (as a function of shifts, order of approximation, 
      and off-centerings) is not easy in the general case. 

      In the particular case of a flat $d+1$-metric 
      with zero shift, the Courant limit is easy to write down:
      $$
      \lambda\leq\frac{\alpha_{0}}{2\sqrt{d C_{n}}}\,,
      $$
      where $C_{n}$ is given in (\ref{boundd2}) and  $\alpha_{0}=R_{lsia}$.
      In the general case, the Courant limit 
      has to be evaluated numerically. 

      For the 1-D wave equation with shift $\beta>0$ 
      with upwind discretization of the advection term
      and adding the minimal amount of dissipation if necessary,
      the limit of the Courant factor is given by 
      \begin{equation}
	\lambda^{(n,s)}(\beta)\equiv\frac{\alpha_{0}}
	       {\underset{\xi\in(\pi,\pi]}{\max}
	       \left\vert
	       \beta
	       \hat{\Omega}^{2(n+1)}
	       \left(
	       \frac{\hat{\mathbf{d}}^{(1,n,s)}}
		    {\hat{\Omega}^{2(n+1)}}-
		    \sigma_{+}^{(n,s)}
		    \right)
		    +i\left(\beta\hat{d}^{(1,n,s)}
		    + \sqrt{\hat{d}^{(2,n)}}\right)
		    \right\vert}\,.
                      \nonumber\end{equation}
                      \begin{figure}[htbp]
                        \centering
    \includegraphics[width=0.325\textwidth,height=0.18\textheight]
		    {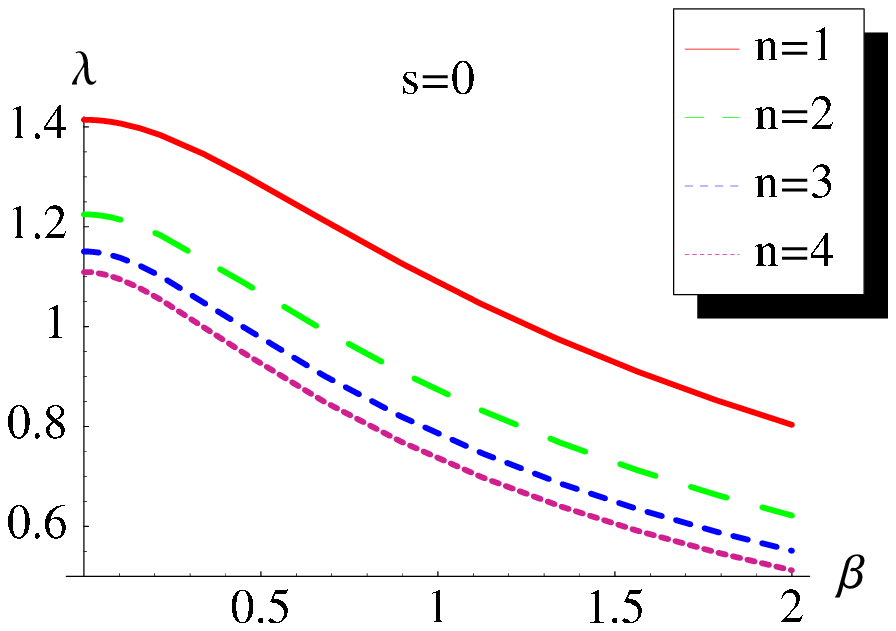}
    \includegraphics[width=0.325\textwidth,height=0.18\textheight]
		    {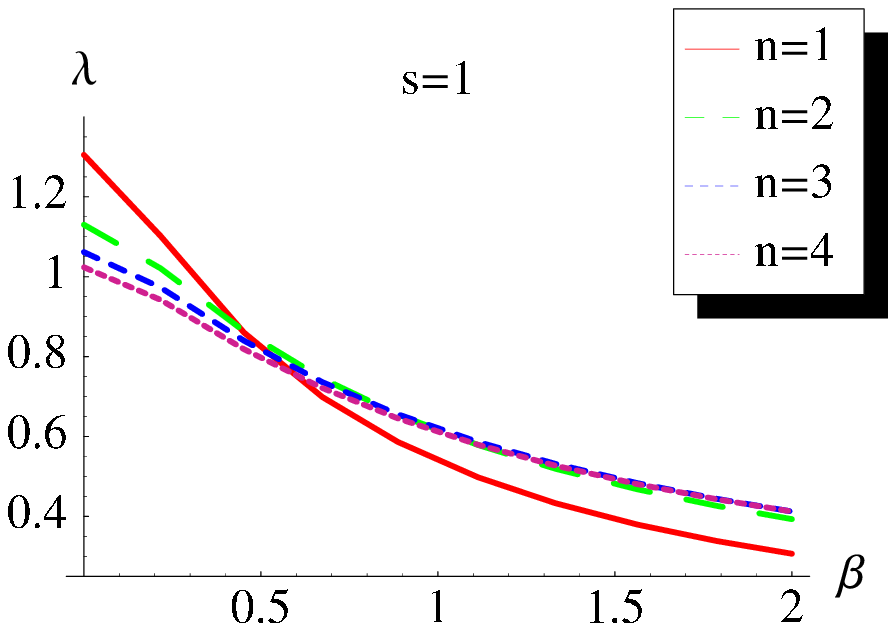}
    \includegraphics[width=0.325\textwidth,height=0.18\textheight]
		    {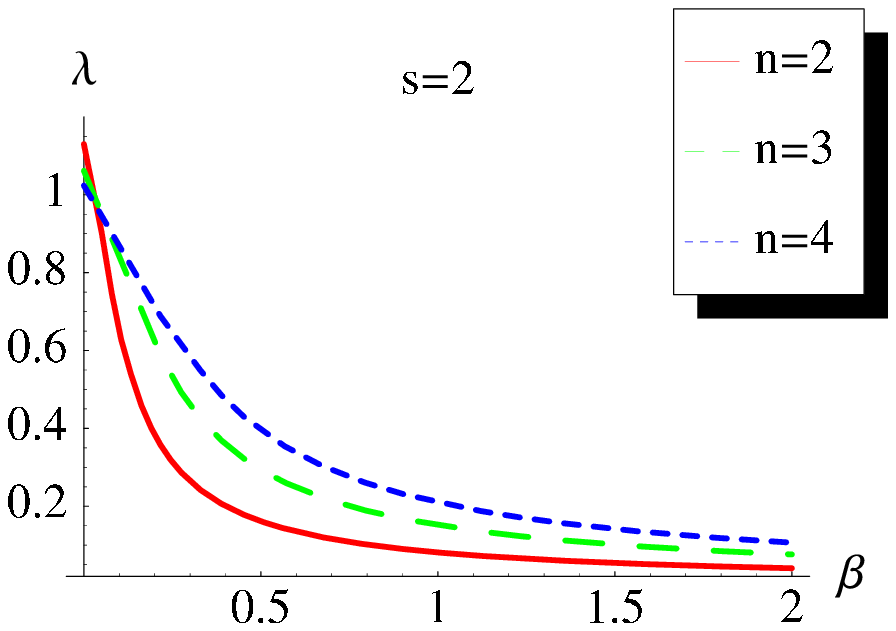}
    \caption{\textbf{Courant limit as a function of $\beta$,
    for different orders of approximation at fixed off-centering $s$}.
    For $s=0$ (left) no dissipation is needed,
    ($\sigma=0$), and we are in the regime of local stability on
    the imaginary axis ($\alpha_{0}=2.83$).
    For $s=1$ (middle),
    again no dissipation is needed ($\sigma=0$),
    but now we are in the regime of local stability ($\alpha_{0}=2.61$).
    For $s=2$ (right) dissipation is required and 
    we add the minimum amount in order to attain stability.
    }
  \label{figCFdiffs}
\end{figure}
\begin{figure}[htbp]
  \centering
    \includegraphics[width=0.325\textwidth,height=0.18\textheight]
		    {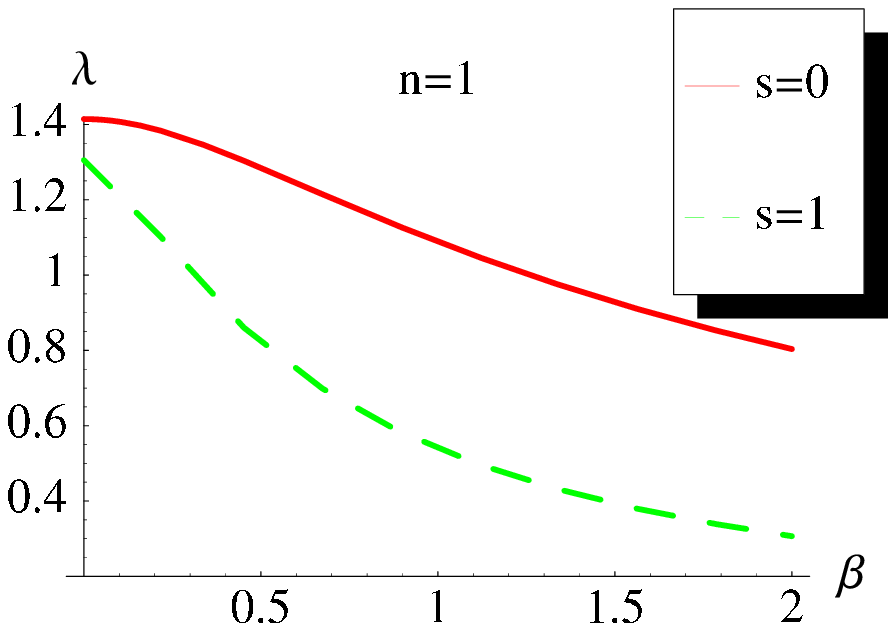}
    \includegraphics[width=0.325\textwidth,height=0.18\textheight]
		    {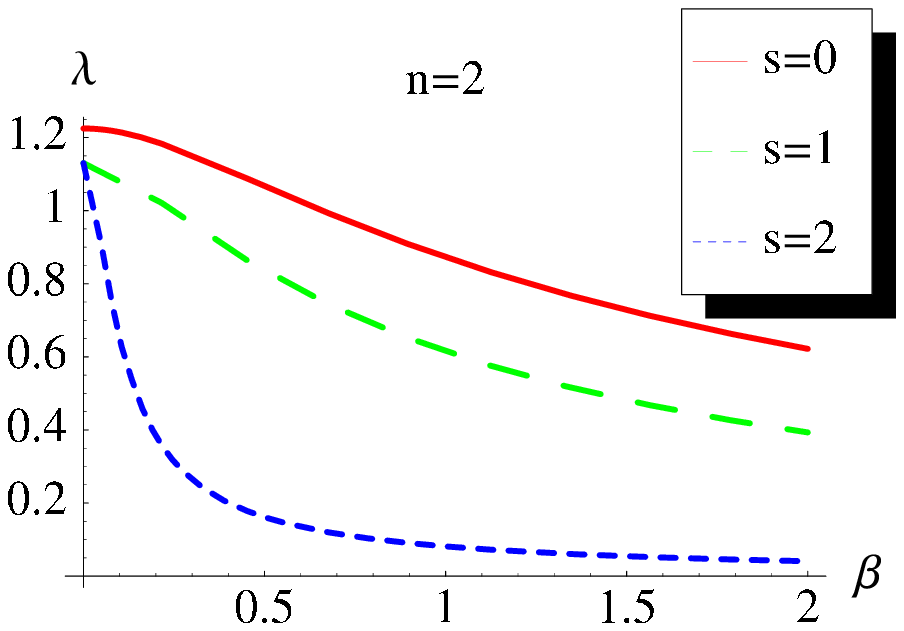}
    \includegraphics[width=0.325\textwidth,height=0.18\textheight]
		    {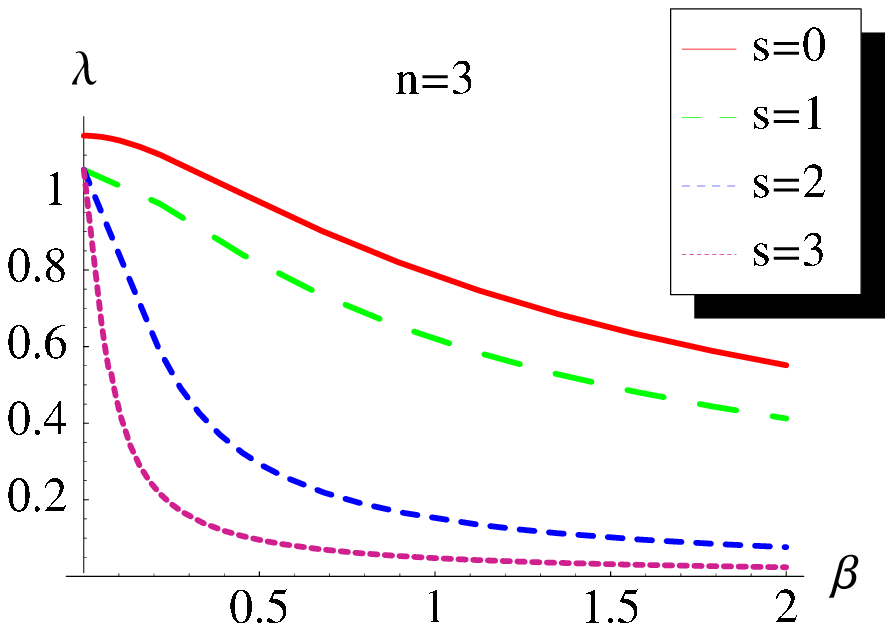}
    \caption{\textbf{Courant limit as a function 
	of $\beta$for different advection stencils 
	at fixed order of spatial accuracy.}
      From left to right: Courant limits at approximation orders $2,4,6$.
      As in Figure \ref{figCFdiffs} the Courant limit calculation
      takes into account whether we are in the regime of
      local stability on the imaginary axis (the case $s=0$),
      or only local stability (for $s\geq 1$),
      and the minimal amount of Kreiss-Oliger dissipation is added for $s\geq 2$.
    }
  \label{figCFdiffn}
\end{figure}
We compare the Courant limits
for different orders of approximations at fixed advection stencil
in Figure \ref{figCFdiffs}, and the Courant limit 
at fixed order of approximation
for different advection stencils in Figure \ref{figCFdiffn}.

In Figure \ref{figCFdiffs} we see that if $s=0$, 
the higher the order of approximation, the 
lower the Courant limit. For $s\geq 1$, this is not true anymore
beyond a certain value of the shift. For large shifts we observe
that increasing the order of approximation, actually {\em decreases}
the Courant limit.

Comparing different stencils in Figure \ref{figCFdiffn}, 
we observe that advecting more 
points decreases the Courant limit,
and there is a significant drop in the Courant factor
between $s=1$ and $s=2$, for all orders of approximation.
\subsection{Phase and Group Speeds}
For the wave equation the continuum phase and group speeds are:
\begin{equation}
  \mathsf{\hat{v}}_{p}=\mathsf{\hat{v}}_{g}
  =\beta^{n}\pm\sqrt{\gamma^{nn}},
  \quad \text{where}
  \quad 
  \mathsf{\hat{v}}_{p}\equiv\frac{\hat{\mathsf{\Lambda}}}
         {\omega_{0}},
  \quad
  \mathsf{\hat{v}}_{g}\equiv n^{j}\frac{d}{d\omega_{j}}
  \hat{\mathsf{\Lambda}}
  \,.
\nonumber\end{equation}
where $\beta^{n}=\beta^{j}n_{j}$ and
$\gamma^{nn}=\gamma^{jl}n_{j}n_{l}$. 
The discrete speeds can be defined in a simillar manner from 
the discrete eigenvalues $\hat{\Lambda}$. 
This would lead to complex speeds in case of 
off-centered schemes which are difficult to investigate \cite{Tref}.
In the following, we assume that the real part of 
the discrete eigenvalues 
has been cancelled by adding appropriate artificial dissipation terms 
(positive or negative). Otherwise, our investigation remains 
valid only at small frequencies, where 
the damping/amplification effect introduced by the real part 
is dominated by the dispersion effect associated 
to the imaginary part of the eigenvalues. 
With these remarks, we define the discrete speeds by:
\begin{eqnarray}
  \hat{v}_{p}&\equiv&  \frac{\hat{\Lambda}^{Im}}{\omega_{0}}
  =\left(\sum_{j=1}^{d}\beta^{j}n_{j}
  \frac{\hat{d}_{j}^{(1,n,s_{j})}}
  {\xi_{j}}\right)
  \pm\frac{(h\hat{\Delta})(\xi)}{\xi_{0}}\,,
  \nonumber\\
  \hat{v}_{g}&\equiv& n^{j}\frac{d}{d\xi_{j}}
  (h\hat{\Lambda}^{Im})
  =\sum_{j=1}^{d}\left[\beta^{j}n_{j}^{2}
  \frac{\partial\hat{d}_{j}^{(1,n,s_{j})}}{\partial\xi_{j}}
  \pm\frac{\partial(h\hat{\Delta})(\xi)}
	  {\partial\xi_{j}}n_{j}\right]\,.
\nonumber
\end{eqnarray}
We also restrict attention to the one dimensional case. 
Because $\pm$ speeds interchange
when $\xi$ changes sign,  it is enough to consider
only the ``+'' speed over the whole spectrum $\xi\in(-\pi,\pi]$.
  Also because we will compare
  speeds at different orders of approximation
  or at different stencils, we attach 
  the superscript $(n,s)$  (or only $(n)$ in case $s=0$), 
  to the symbols representing 
  the discrete speeds and the corresponding errors:
\begin{eqnarray}
  \hat{v}^{(n,s)}_{p}(\xi)&=&\frac{1}{\xi}\left(\beta\hat{d}^{(1,n,s)}
    +\sqrt{\hat{d}^{(2,n)}}\right)\,,\nonumber\\
  \hat{v}^{(n,s)}_{g}(\xi)&=&\frac{d}{d\xi}\left(\beta\hat{d}^{(1,n,s)}
  +\sqrt{\hat{d}^{(2,n)}}\right)\,.\nonumber
\end{eqnarray}
The continuum limits for both, phase and group speeds 
are $\beta+1$ for $\xi>0$
and $\beta-1$ for $\xi<0$.
We will analyze the behavior of the speed errors defined as
\begin{eqnarray}
  \hat{\epsilon}^{(n,s)}_{p}&\equiv&
  \beta\left(\frac{\hat{d}^{(1,n,s)}}{\xi}- 1\right)
  +\left(\frac{\sqrt{\hat{d}^{(2,n)}}}{\xi}-\sign{\xi}\right)\,,
  \label{PhaseErrDef}
  \\
  \hat{\epsilon}^{(n,s)}_{g}&\equiv&
  \beta\left(\frac{d}{d\xi}\hat{d}^{(1,n,s)}-1\right)
  +\left(\frac{d}{d\xi}\sqrt{\hat{d}^{(2,n)}}-\sign{\xi}\right)\,.
  \label{GroupErrDef}
\end{eqnarray}
We will also assume $\beta\geq 0$ without restricting generality,
if $\beta\rightarrow -\beta$, then 
$\hat{\epsilon}^{(n,s)}_{p,g}(\xi)\rightarrow -\hat{\epsilon}^{(n,s)}_{p,g}(-\xi)$.
\subsubsection{Small Frequencies}
When $\xi\simeq 0$ one can show that the phase and group speed
errors satisfy
\begin{eqnarray}
  \hat{\epsilon}^{(n,s)}_{p}&=&
  -\left\vert c_{n}\right\vert\left[
    (-1)^{s}\frac{(n+s)!(n-s)!}{(n!)^2}
    \beta+\frac{\sign{\xi}}{2(n+1)}
    \right]\xi^{2n}+O(\xi^{2n+2})\,,\nonumber\\
  \hat{\epsilon}^{(n,s)}_{g}&=&
  -(2n+1)\left\vert c_{n}\right\vert\left[
    (-1)^{s}\frac{(n+s)!(n-s)!}{(n!)^2}
    \beta+\frac{\sign{\xi}}{2(n+1)}
    \right]\xi^{2n}+O(\xi^{2n+2})\,.\nonumber
\end{eqnarray}
Because the errors scale  with $\xi^{2n}$, 
it is obvious that for small enough frequencies higher order 
approximations will improve the phase and group errors for all the
values of the shift and for all advection stencils.\\
If we keep the order fixed and compare the speeds corresponding 
to an off-centering by $s\geq 1$-points with the ones corresponding to 
the centered scheme, $s=0$, then one can easily show that the 
off-centered scheme improves over the centered  one
\begin{itemize}
\item the ``+'' numerical speeds ($\xi>0$) if $s$ is odd and $\beta$
  is small enough
\item  the ``-'' numerical speeds ($\xi<0$) if $s$ is even 
  and $\beta$ is small enough
\end{itemize}
where small enough means 
\begin{equation}
\beta<\frac{1}{(n+1)}\frac{1}{\frac{(n+s)!(n-s)!}{(n!)^2}-1}.
\nonumber
\end{equation}
\textbf{Obs.} For $s=1$, the above inequality becomes $\beta<\frac{n}{(n+1)}$.
Also notice that with increasing $s$ the  above limit on $\beta$ decreases.\\
In the next subsection we will analyze the behavior
for the whole spectrum in some more detail.

\subsubsection{Comparison with wave equation written in first order form} 
If the wave equation is written 
in first order form (approximating the first derivatives
with the corresponding CFDO), then the eigenvalues become
$(h\hat{\Lambda}_{\pm})(\xi)=i\left(\beta\pm 1\right)\hat{d}^{(1,n)}$.
For $\xi\simeq 0$ one then gets
\begin{eqnarray}
  \hat{\epsilon}^{(n)}_{p}&=&-\left(\beta+\sign{\xi}\right)
  \left\vert c_{n}\right\vert
    \xi^{2n}+O(\xi^{2n+2})\nonumber\\
    \hat{\epsilon}^{(n)}_{g}&=&-\left(\beta +\sign{\xi}\right)
    (2n+1)\left\vert c_{n}\right\vert
    \xi^{2n}+O(\xi^{2n+2}).\nonumber
\end{eqnarray}
We notice that for a given order, the second order system 
discretized with CFDO, has smaller phase and group errors 
then the first order one (for both eigenvalues), 
if and only if $\left\vert\beta\right\vert\leq\frac{2n+3}{4(n+1)}$.
If $\left\vert\beta\right\vert$ is not in this interval then 
one pair of speeds (phase and group) is better approximated 
by the second order system,
while the other one is better approximated
by the first order system.

\subsubsection{Scaling of the Speed Errors
  with the Order of Approximation}
\begin{figure}[htbp]
  \centering
  \includegraphics[width=0.45\textwidth]{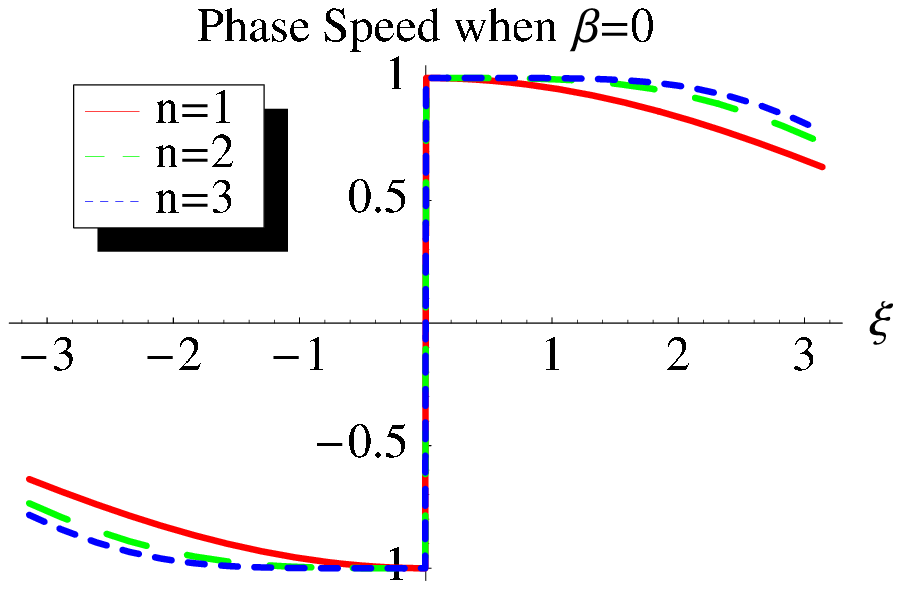}
  \includegraphics[width=0.45\textwidth]{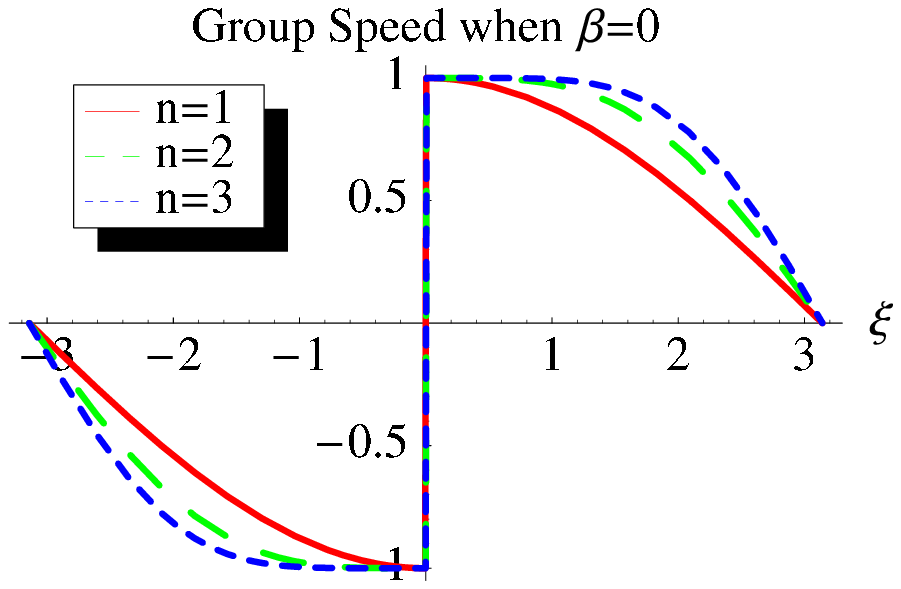}
  \caption{\textbf{Phase and Group Speeds for $\beta=0$.}
    The higher the order of the approximation, the more accurate
    the phase and group speeds are for all frequencies.}
  \label{figbeta0}
\end{figure}

\begin{lemma}
  If $\beta=0$, then increasing the order of approximation decreases the
  phase and group speed errors for all frequencies.
\end{lemma}
To prove this we make use of the relations (\ref{derivsymbols})
in the definitions of the speeds and obtain
\begin{eqnarray}
  \hat{v}^{(n)}_{p}=
  \frac{\sqrt{\hat{d}^{(2,n)}}}{\xi},
  &&  \hat{\epsilon}^{(n)}_{p}=
  \frac{\sqrt{\hat{d}^{(2,n)}}}{\xi}-\sign{\xi}\,,\nonumber
  \\\nonumber\\
  \hat{v}^{(n)}_{g}=
  \frac{\hat{d}^{(1,n)}}{\sqrt{\hat{d}^{(2,n)}}},
  &&\hat{\epsilon}^{(n)}_{g}=
  \frac{\hat{d}^{(1,n)}}{\sqrt{\hat{d}^{(2,n)}}}-\sign{\xi}\,.\nonumber
\end{eqnarray}
Using the inequalities (\ref{ineqd2}) and 
(\ref{ineqr}) one can easily show  that
$\left\vert\hat{\epsilon}^{(n+1)}_{p}\right\vert<
\left\vert\hat{\epsilon}^{(n)}_{p}\right\vert$
and  $\left\vert\hat{\epsilon}^{(n+1)}_{g}\right\vert<
\left\vert\hat{\epsilon}^{(n)}_{g}\right\vert$ for all frequencies.
The situation is illustrated in Figure \ref{figbeta0} where 
we plot the speeds $\hat{v}^{(n)}_{p}$ and  $\hat{v}^{(n)}_{g}$ versus $\xi$.

\begin{figure}[htbp]
  \centering
  \begin{minipage}[b]{13 cm}
  \includegraphics[width=0.325\textwidth,height=0.18\textheight]{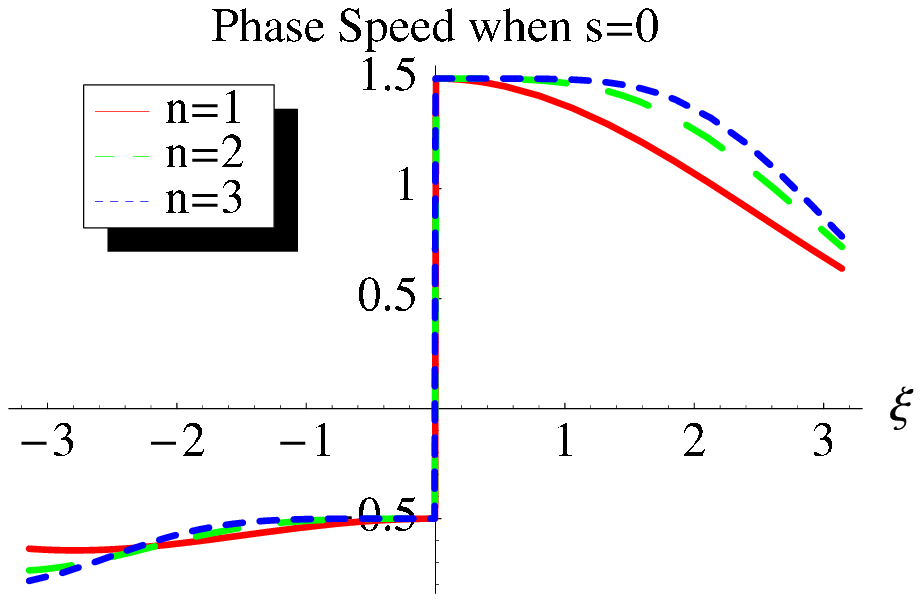}
  \includegraphics[width=0.325\textwidth,height=0.18\textheight]{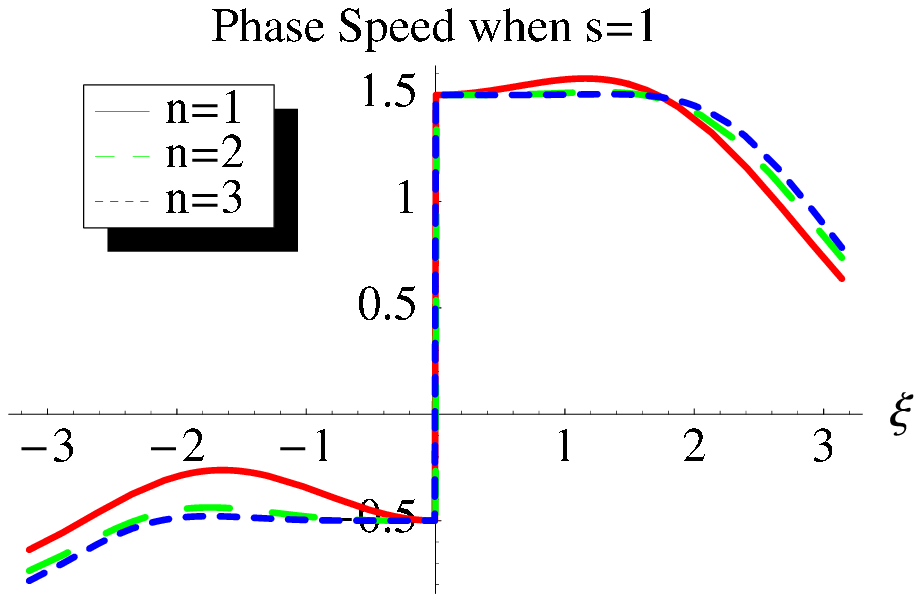}
  \includegraphics[width=0.325\textwidth,height=0.18\textheight]{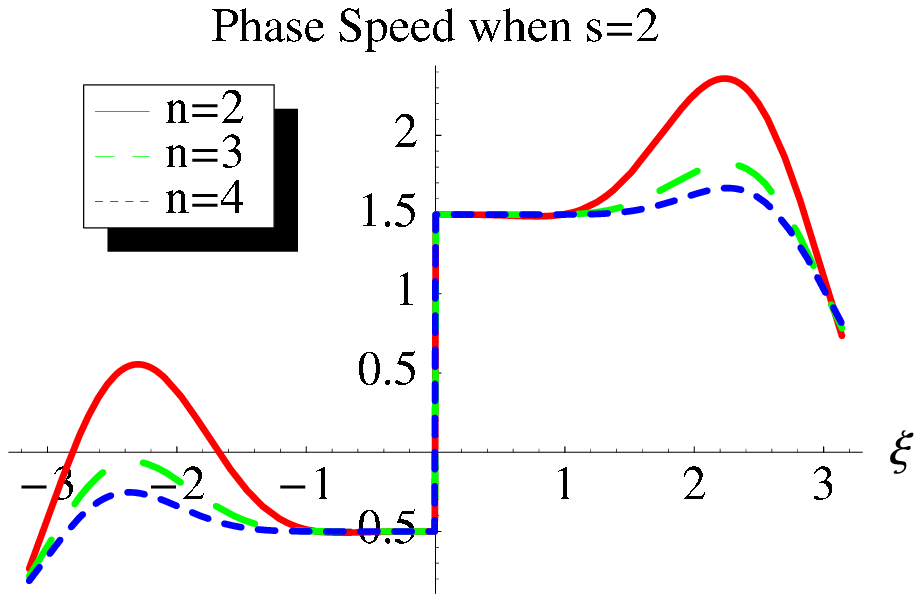}
  \end{minipage}
  \begin{minipage}[b]{13 cm}
  \includegraphics[width=0.325\textwidth,height=0.18\textheight]{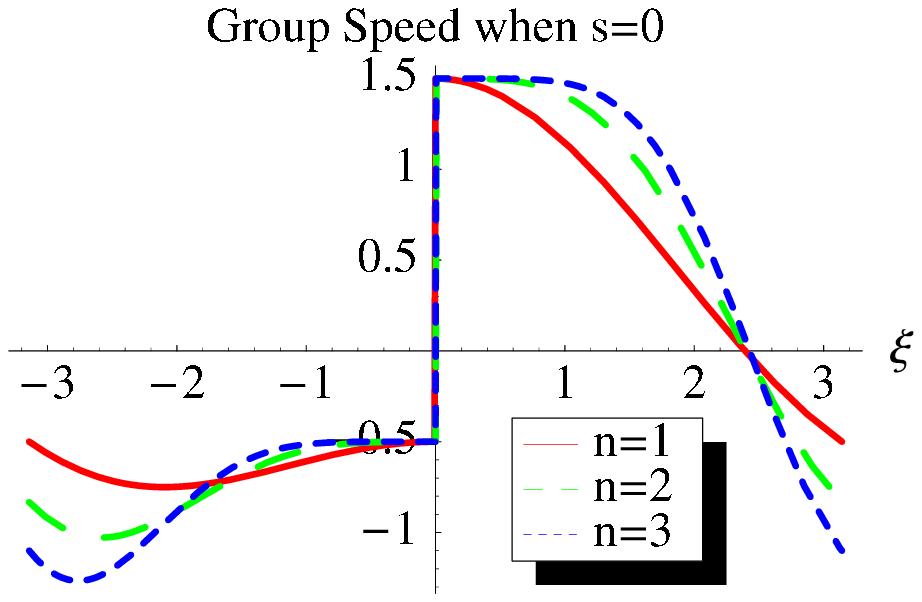}
  \includegraphics[width=0.325\textwidth,height=0.18\textheight]{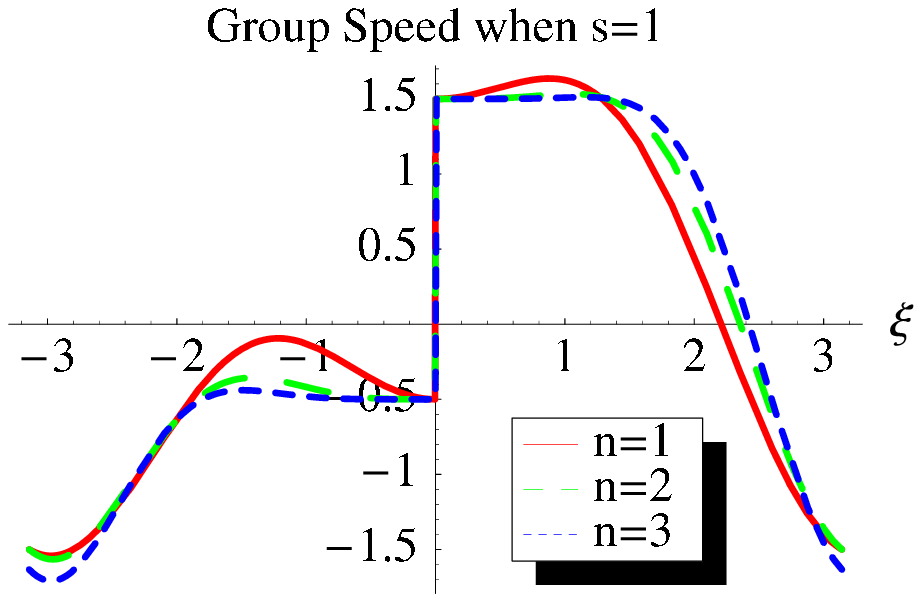}
  \includegraphics[width=0.325\textwidth,height=0.18\textheight]{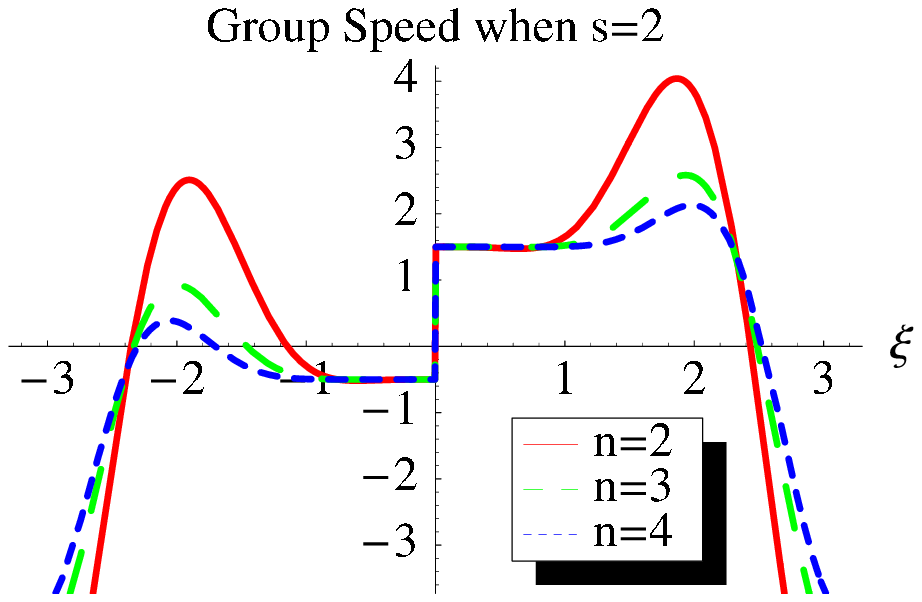}
  \end{minipage}
  \caption{The phase and group speeds at different orders  
     of approximation but keeping the level of off-centering fixed, for
     $\beta=0.5$.}
    \label{figpnPhaseGroupErrors}
\end{figure}

If $\beta\neq0$ then  it is not true anymore 
that higher order approximations improve the numerical speeds
for all frequencies (not even for the case of using CFDO).
Though one can go into details and determine the regions
in the spectrum where the scaling with order fails,
we restrict ourselves to illustrating this situation
by plotting the numerical speeds versus frequency
at a particular value of the shift.
In Figure \ref{figpnPhaseGroupErrors}
we show the numerical speeds at different orders
of approximation with the same advection stencil when $\beta=0.5$.
\subsubsection{Scaling of the Speed Errors with Off-centering}
\label{SpeedsFixedOrder}
\begin{figure}[htbp]
  \centering
  \includegraphics[width=0.325\textwidth,height=0.18\textheight]{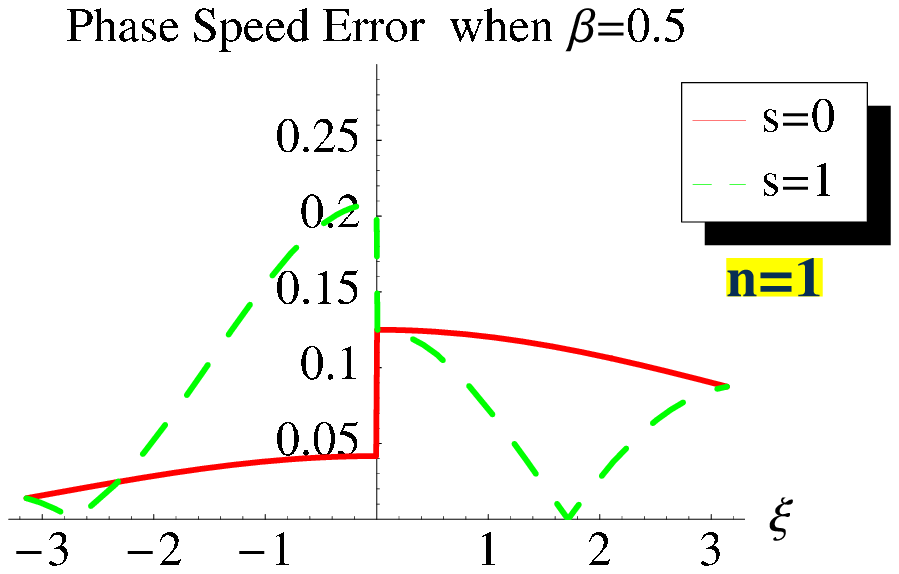}
  \includegraphics[width=0.325\textwidth,height=0.18\textheight]{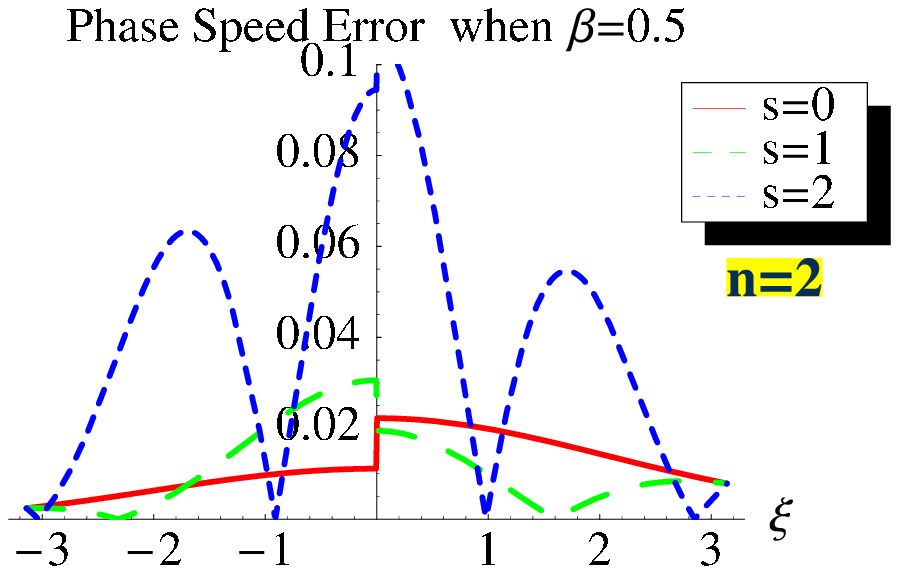}
  \includegraphics[width=0.325\textwidth,height=0.18\textheight]{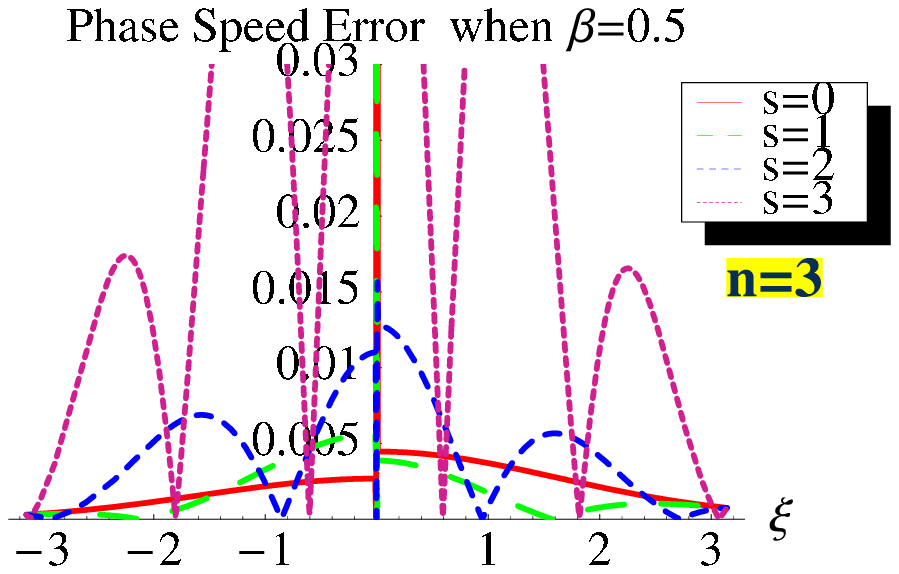}
  \includegraphics[width=0.325\textwidth,height=0.18\textheight]{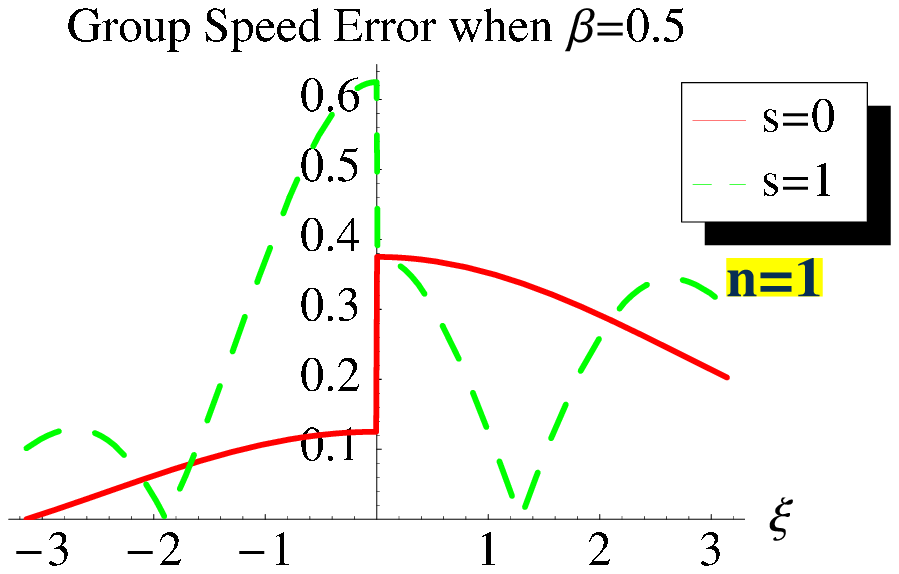}
  \includegraphics[width=0.325\textwidth,height=0.18\textheight]{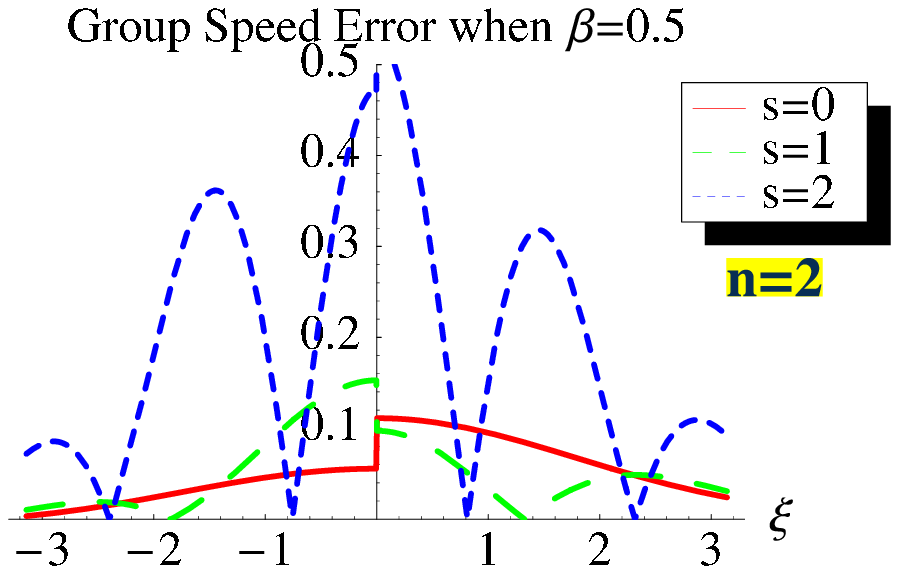}
  \includegraphics[width=0.325\textwidth,height=0.18\textheight]{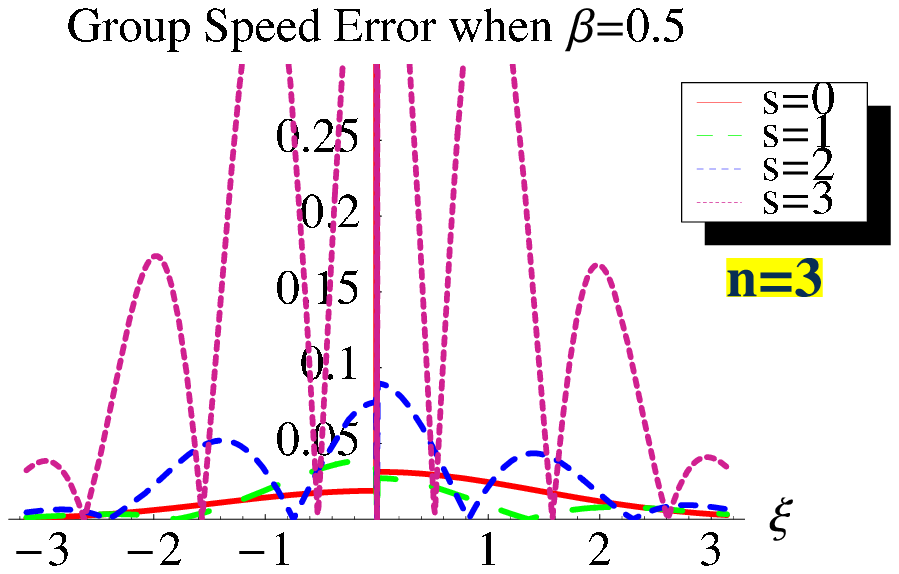}
   \caption{The phase and group speeds errors (scaled with $\xi^{2n}$)
     are shown 
    at different advection stencils for $\beta=0.5$.} 
    \label{figPhaseGroupErrorsFixedOrder}
\end{figure}

The next question we want to answer is what happens 
with the numerical speed errors, 
if we keep the order of approximation fixed and 
vary the off-centering of the first derivative.
For example, we illustrate this situation 
in Figure \ref{figPhaseGroupErrorsFixedOrder}, when  $\beta=0.5$.
As one can see already in these plots, although we know that  
off-centering increases the error of 
the finite difference operator, it is not necessary 
that the numerical speeds will follow the same pattern.
E.g. for this value of the shift, 
the "+'' speed seems more accurate with $s=1$ 
than with $s=0$ for all the spectrum.
In the following we will determine the regions in the 
$(\xi,\beta)$-plane where off-centering improves 
the numerical speed errors over the centered scheme.
\subsubsection*{Phase Speeds}
\begin{figure}[th]
  \centering
  \includegraphics[width=0.325\textwidth,height=0.2\textheight]
		  {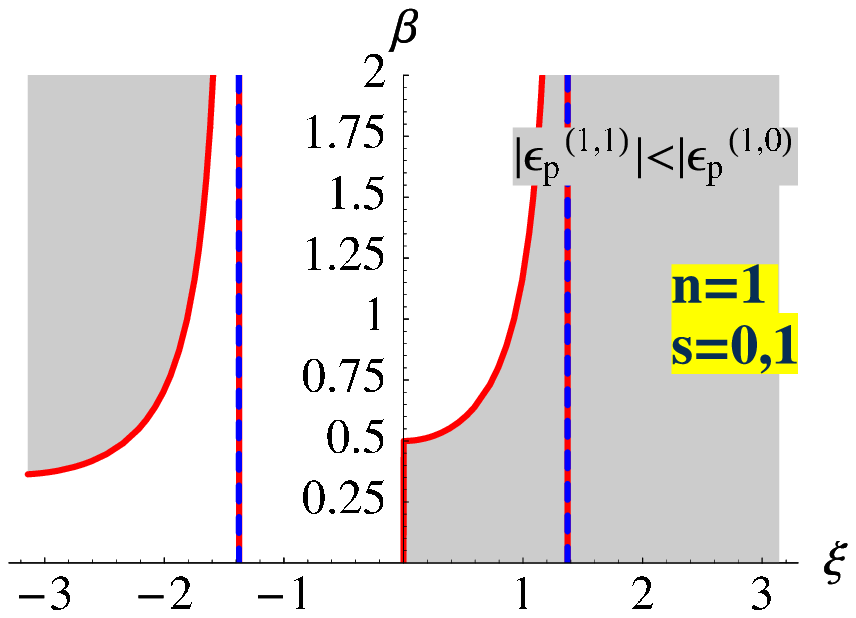}
  \includegraphics[width=0.325\textwidth,height=0.2\textheight]
		  {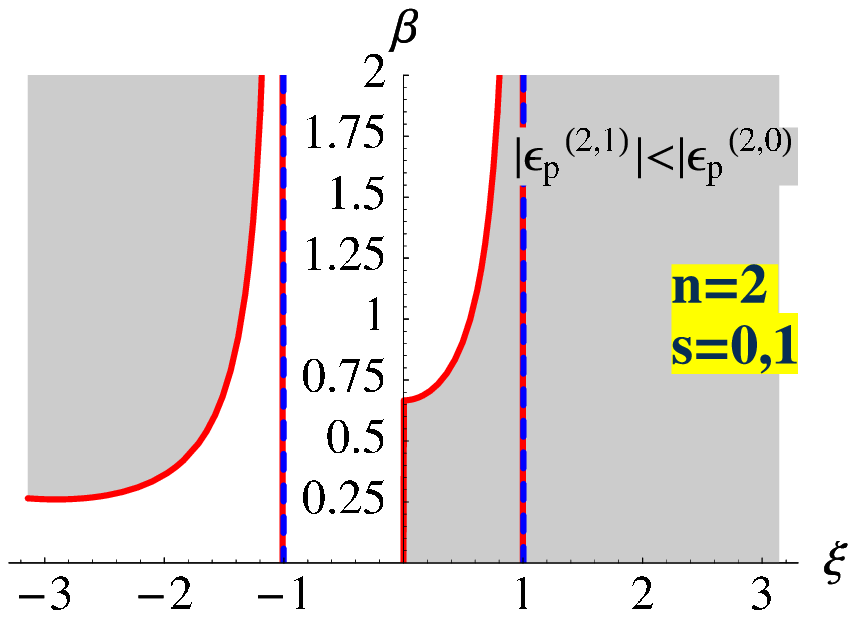}
  \includegraphics[width=0.325\textwidth,height=0.2\textheight]
		  {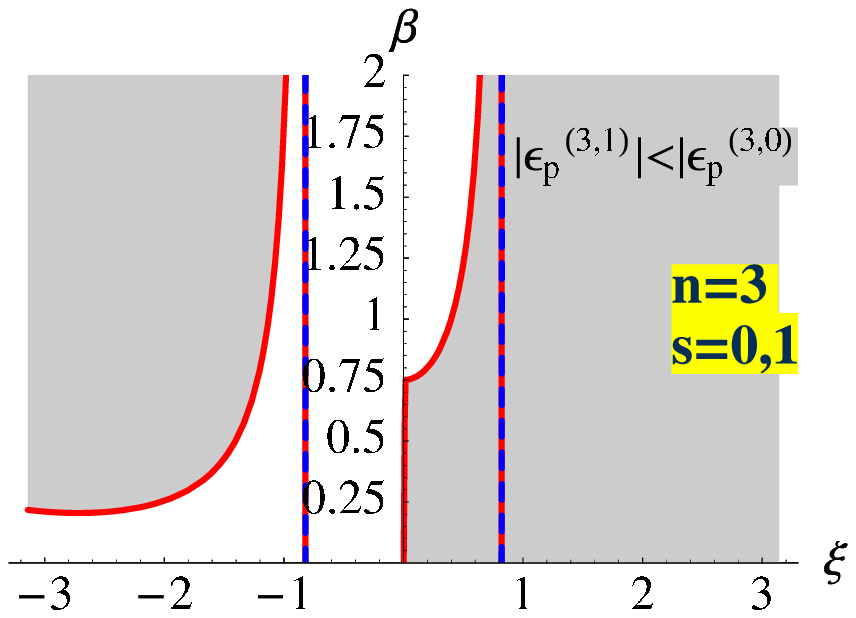}
  \includegraphics[width=0.325\textwidth,height=0.2\textheight]
		  {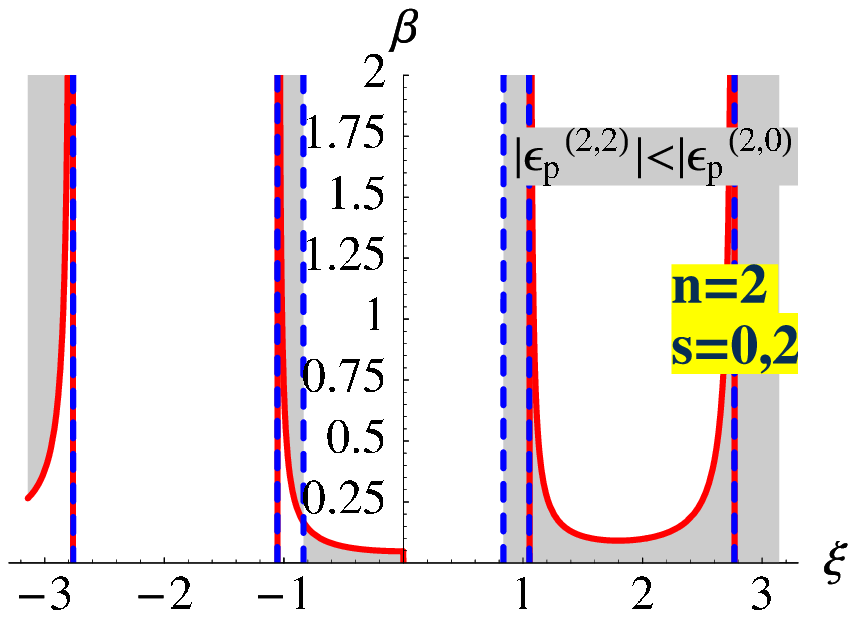}
  \includegraphics[width=0.325\textwidth,height=0.2\textheight]
		  {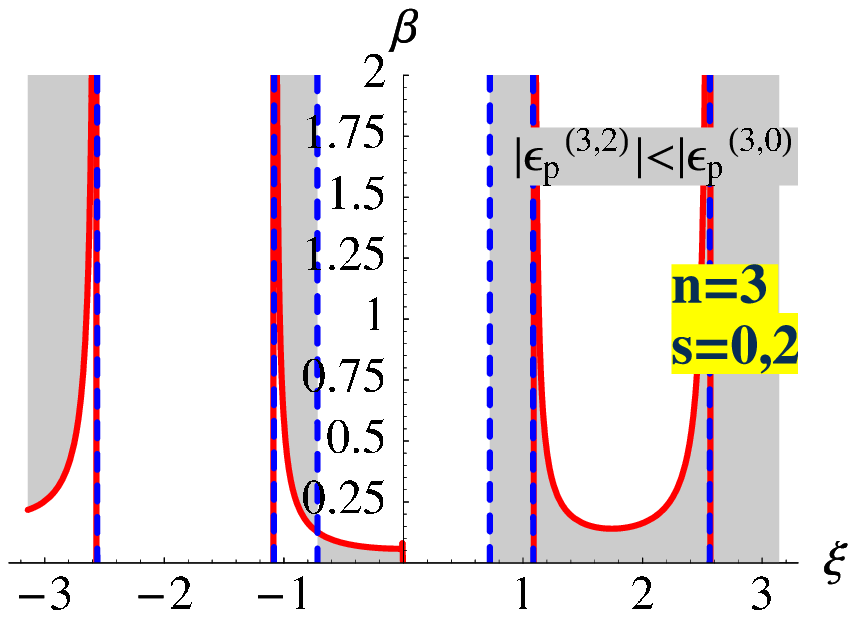}
  \includegraphics[width=0.325\textwidth,height=0.2\textheight]
		  {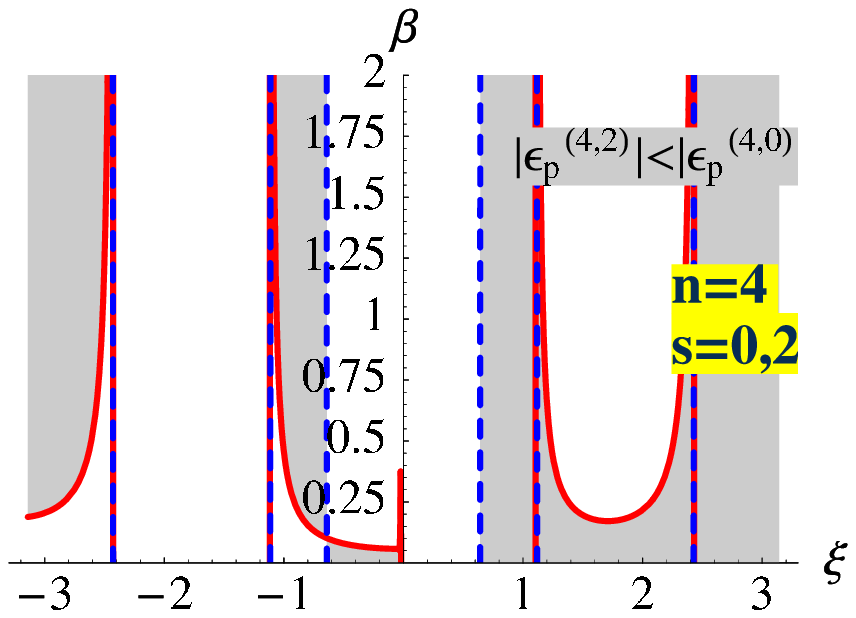}
  \includegraphics[width=0.325\textwidth,height=0.2\textheight]
		  {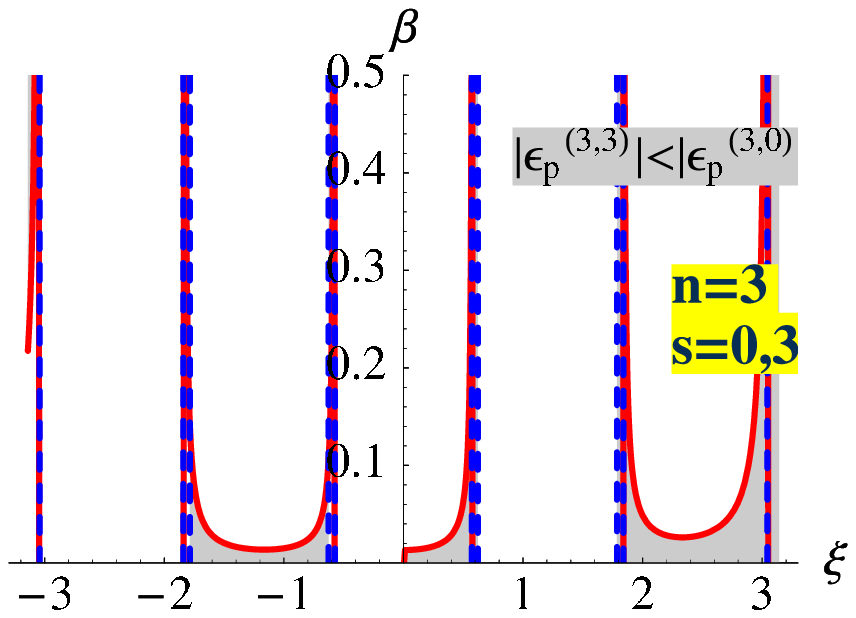}
  \includegraphics[width=0.325\textwidth,height=0.2\textheight]
		  {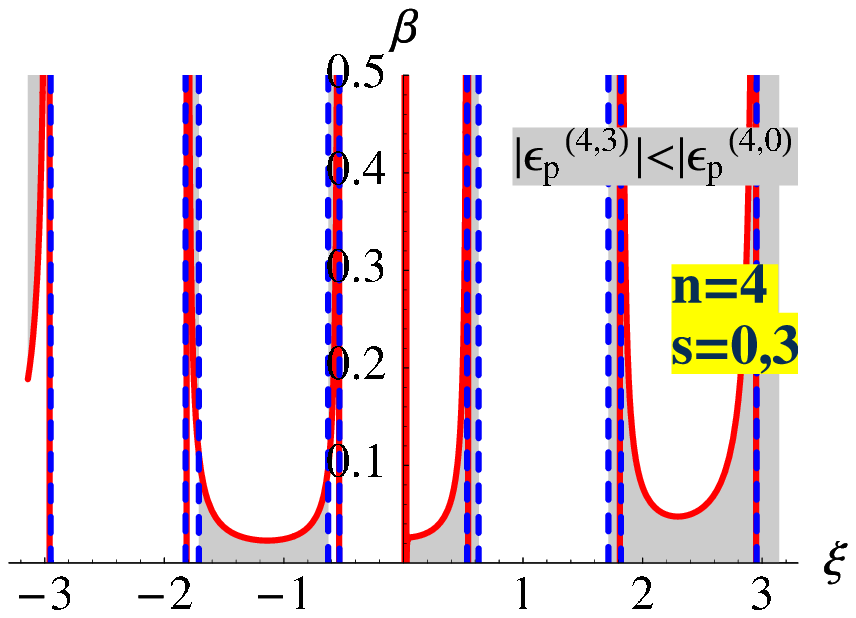}
  \includegraphics[width=0.325\textwidth,height=0.2\textheight]
		  {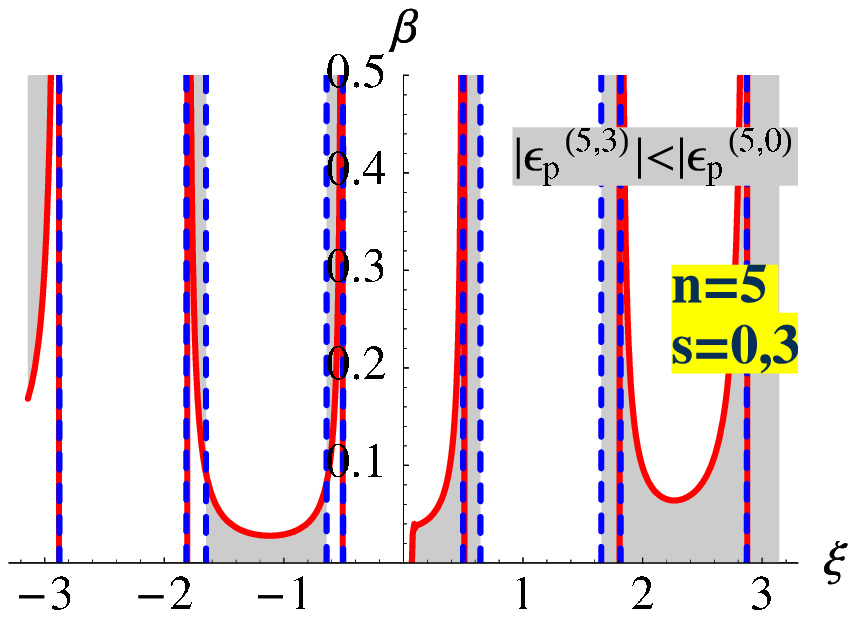}
  \caption{
    Shown are the regions where
    advected stencils improve the phase speed error over
    the centered scheme. The regions are delimited
    by the quantity $\beta^{(n,s)}$ and the zeros of the function
    $f_{1}^{(n,s)}$ as defined in (\ref{funcPhaseSt}).}
  \label{RegPhaseFixedOrder}
\end{figure}

Imposing $\left\vert\hat{\epsilon}^{(n,s)}_{p\pm}\right\vert<
\left\vert\hat{\epsilon}^{(n,0)}_{p\pm}\right\vert$ and using
the definition (\ref{PhaseErrDef})
yields the inequality
\begin{equation}
  f^{(n,s)}_{1}(\xi) f^{(n,s)}_{2}(\xi)\left(\beta
  -\beta^{(n,s)}(\xi)\right)<0\,,
  \label{ineqPhSpeed}
\end{equation}
where
\begin{eqnarray}
  f^{(n,s)}_{1}(\xi)&\equiv&\hat{d}^{(1,n,s)}-\hat{d}^{(1,n,0)}\,,
  \nonumber\\
  f^{(n,s)}_{2}(\xi)&\equiv&\hat{d}^{(1,n,s)}+\hat{d}^{(1,n,0)}-2\xi\,,
  \nonumber\\
  g^{(n)}(\xi)&\equiv&2\left(\left\vert\xi\right\vert
  -\sqrt{\hat{d}^{(2,n)}}\right)\,,\nonumber\\
  \beta^{(n,s)}(\xi)&\equiv&\frac{g^{(n)}(\xi)}{f^{(n,s)}_{2}(\xi)}\,.
\label{funcPhaseSt}
\end{eqnarray}
We have $g^{(n)}(\xi)>0$ but $f^{(n,s)}_{1,2}$ can change 
sign over the spectrum.
The inequality (\ref{ineqPhSpeed}) holds at a given frequency $\xi$, 
if $\beta>\beta^{(n,s)}(\xi)$ and $\sign{f^{(n,s)}_{1,2}}(\xi)<0$
or $\beta<\beta^{(n,s)}(\xi)$ and $\sign{f^{(n,s)}_{1,2}}(\xi)>0$.
In general, the regions in $(\xi,\beta)$ plane where 
at fixed order of approximation, 
off-centering by $s$ points improves the accuracy of the phase speed, 
are difficult to determine analytically and we restrict ourselves 
to a numerical evaluation (Figure \ref{RegPhaseFixedOrder}).
What we see in the plots is that if $s$ is odd (even) 
then for sufficiently small $\beta$,
the ``+'' (``-'') speed has smaller error 
compared with the case of CFDO in some intervals of the spectrum 
that include the small frequency range.
However these regions become narrower with increasing the off-centering, 
such that for $s=1$ we have the strongest effect. 
We analyze this case in  more detail below.\\
If \textbf{$s=1$}, the functions $f^{(n,s)}_{1,2}$ and $\beta^{(n,s)}$, 
  defined in (\ref{funcPhaseSt}) become
\begin{eqnarray}
  f^{(n,1)}_{1}(\xi)&=& \frac{\left\vert c_{n-1}\right\vert}{2}
  (\sin\xi)
  \hat{\Omega}^{2 n}\,,\nonumber\\ 
  f^{(n,1)}_{2}(\xi)&=& \hat{\delta}\frac{\left\vert c_{n-1}\right\vert}{2}
  \hat{\Omega}^{2 n}
  +2\left(\hat{d}^{(1,n)}-\xi\right)\,,\nonumber\\
  \beta^{(n,1)}(\xi)&=&\frac{2\left(\left\vert\xi\right\vert
  -\sqrt{\hat{d}^{(2,n)}}\right)}{f^{(n,1)}_{2}(\xi)}\,.
\end{eqnarray}
We have $\sign{f^{(n,1)}_{1}(\xi)}=\sign{\xi}$
and $f^{(n,1)}_{1}(\pm\pi)=0$.\\
Then the inequality $\left\vert\hat{\epsilon}^{(n,1)}_{p}\right\vert<
\left\vert\hat{\epsilon}^{(n,0)}_{p}\right\vert$ holds
\begin{itemize}
\item for $\xi>0$ if $\beta^{(n,1)}(\xi)<0$ or $0<\beta<\beta^{(n,1)}(\xi)$,
\item for $\xi<0$ if $\beta>\beta^{(n,1)}(\xi)>0$.
\end{itemize}
The limits of $\beta^{(n,1)}(\xi)$ in $0$ and $\pi$ are
\begin{eqnarray}
  \underset{\xi\searrow 0}{\lim}\beta^{(n,1)}(\xi)
  &=&-\underset{\xi\nearrow 0}{\lim}\beta^{(n,1)}(\xi)=\frac{n}{n+1}\,,
  \nonumber\\
  \beta^{(n,1)}(\pi)&=&-1+2\frac{\sqrt{C_{n}}}{\pi}<0\,.
  \nonumber
\end{eqnarray}
It can be shown that the equation $\beta=\beta^{(n,1)}(\xi)$ 
has at most one solution in each of the branches
$\xi>0$ and $\xi<0$, that we will denote by $\xi^{\pm}$.
It turns out that $\left\vert\hat{\epsilon}^{(n,1)}_{p}\right\vert<
\left\vert\hat{\epsilon}^{(n,0)}_{p}\right\vert$ holds if
\begin{itemize}
\item $\beta<1-2\frac{\sqrt{C_{n}}}{\pi}$ and
  $\xi\in (0,\pi)$,
\item $1-2\frac{\sqrt{C_{n}}}{\pi}<\beta<\frac{n}{n+1}$ 
  and $\xi\in(-\pi,\xi^{-})\cup(0,\pi)$,
\item $\beta>\frac{n}{n+1}$ and 
 $\xi\in(-\pi,\xi^{-})\cup(\xi^{+},\pi)$.
\end{itemize}
At a given order of approximation, $2n$, 
for sufficiently small $\beta<\frac{n}{n+1}$
the ``+'' speed has smaller error 
in the case when we advect one point than in the case
when we use CFDO, for all frequencies $0<\xi\leq\pi$, but
the ``-'' speed will have larger error, at least 
for small and mid frequencies.
 
If $\beta>\frac{n}{n+1}$ then for both $\pm$ speeds,
in the regime of small frequencies,
the CFDO give less error than one-point advected 
scheme, while for mid and high frequencies
the situation reverses.
The interval of small frequencies where CFDO are better
than advected scheme shrinks with increasing the order
of approximation.

\subsubsection*{Group Speeds}
\begin{figure}[th]
  \centering
  \includegraphics[width=0.325\textwidth,height=0.2\textheight]
		  {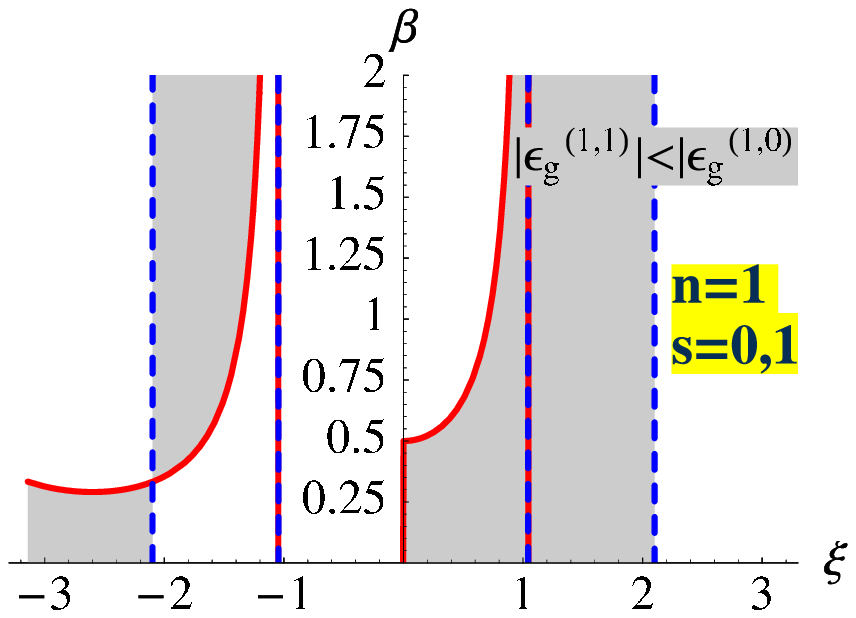}
  \includegraphics[width=0.325\textwidth,height=0.2\textheight]
		  {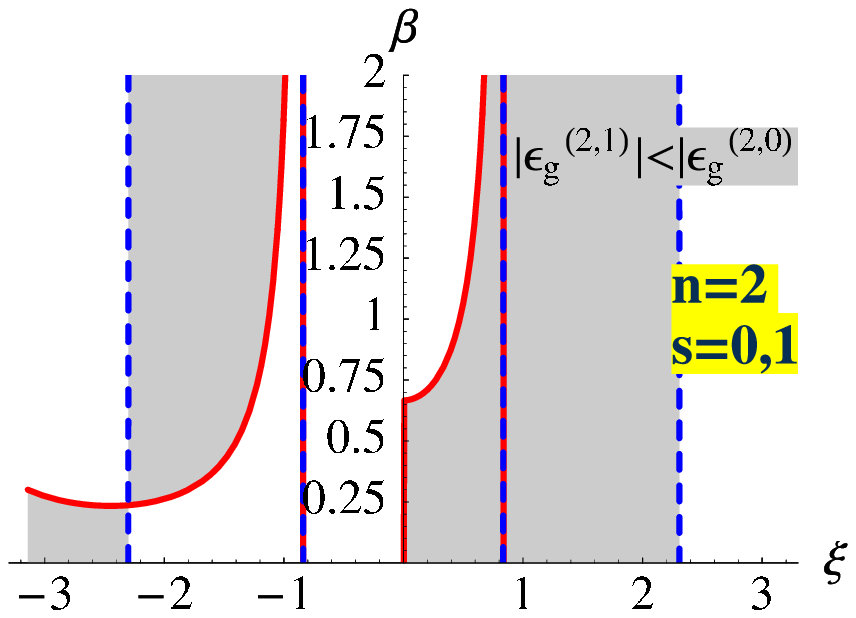}
  \includegraphics[width=0.325\textwidth,height=0.2\textheight]
		  {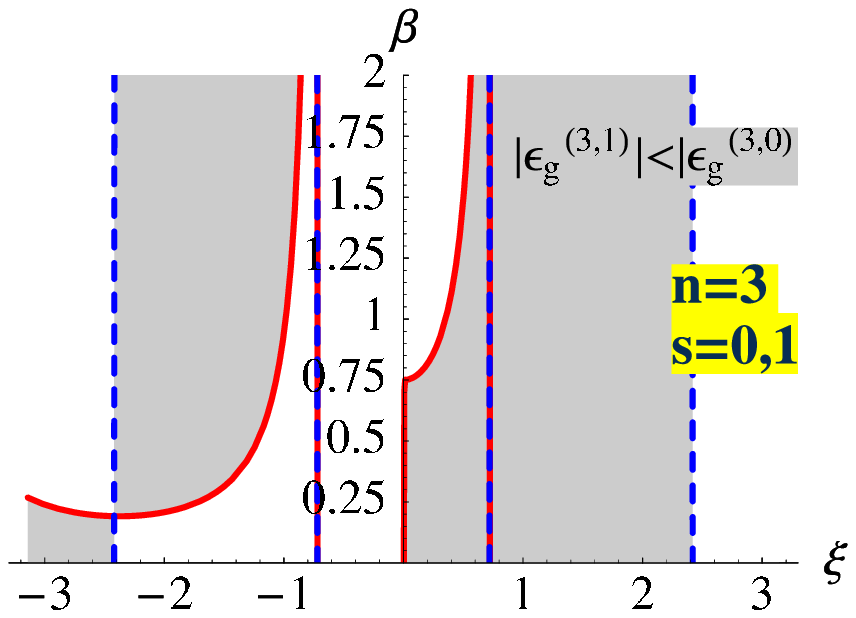}
  \includegraphics[width=0.325\textwidth,height=0.2\textheight]
		  {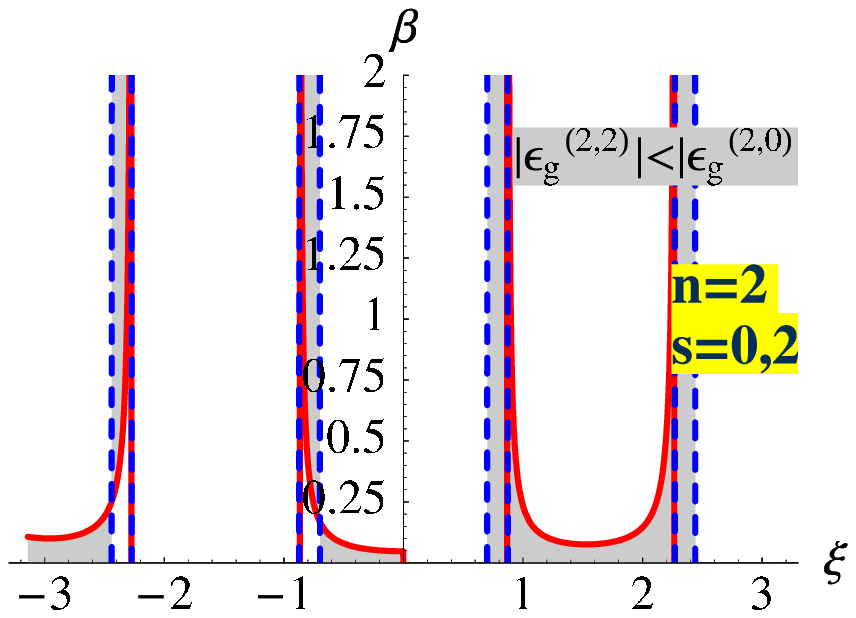}
  \includegraphics[width=0.325\textwidth,height=0.2\textheight]
		  {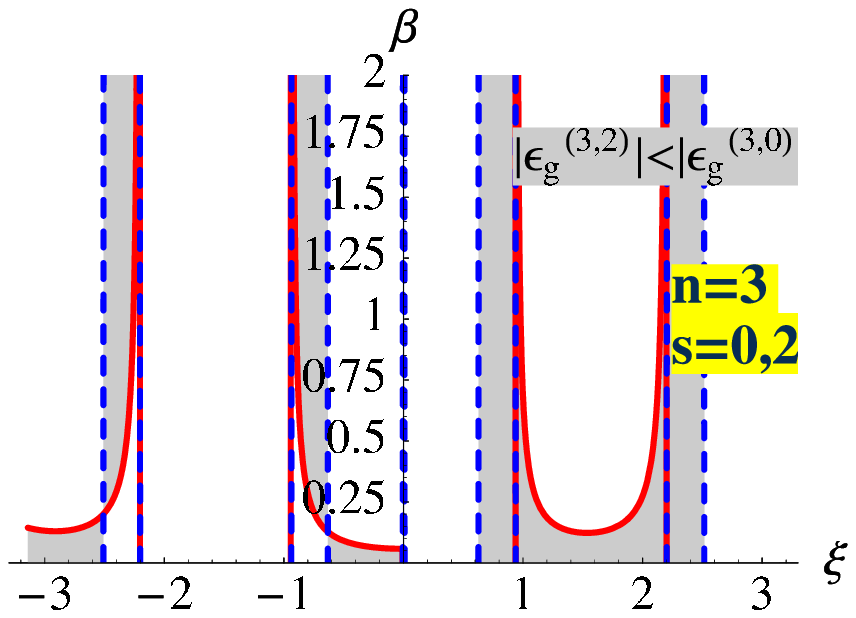}
  \includegraphics[width=0.325\textwidth,height=0.2\textheight]
		  {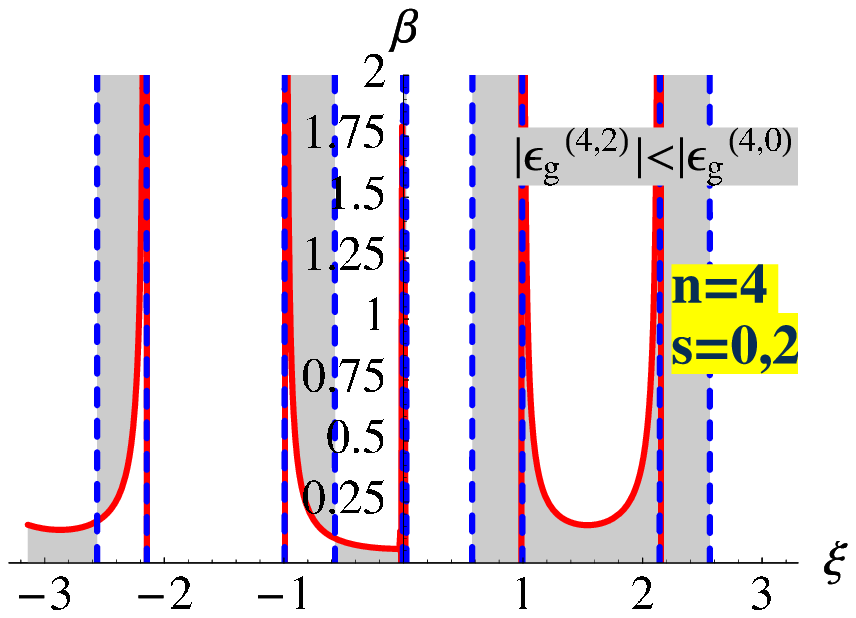}
  \includegraphics[width=0.325\textwidth,height=0.2\textheight]
		  {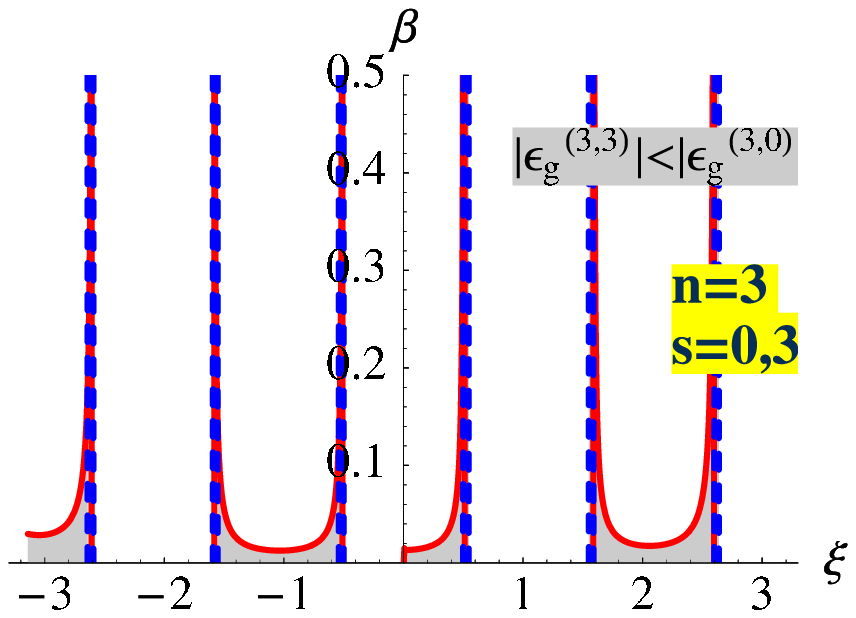}
  \includegraphics[width=0.325\textwidth,height=0.2\textheight]
		  {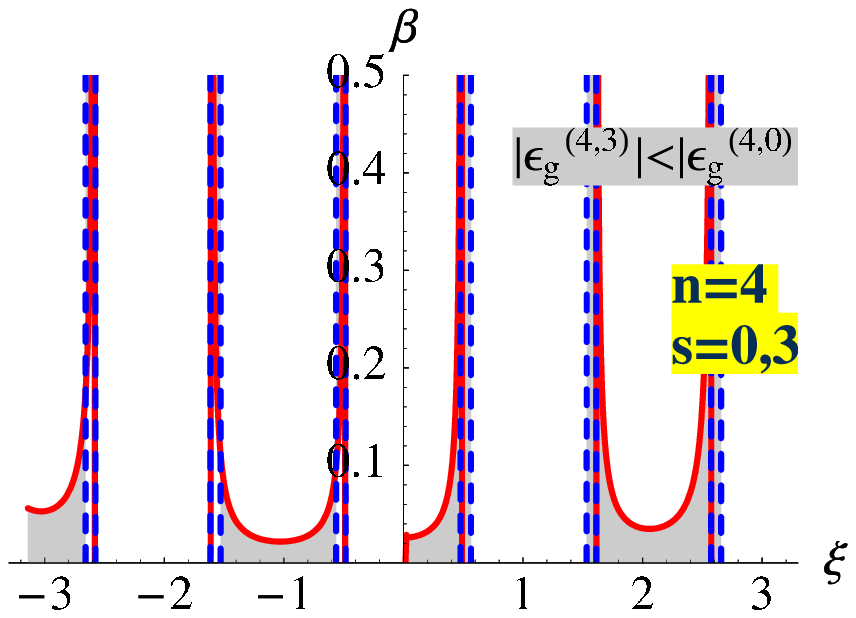}
  \includegraphics[width=0.325\textwidth,height=0.2\textheight]
		  {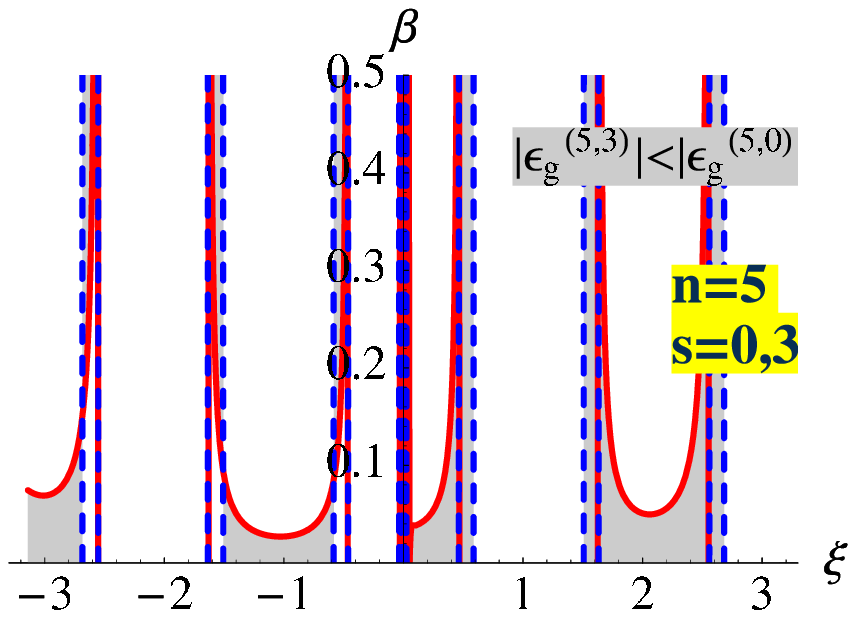}
  \caption{
    Shown are the regions where
    advected stencils improve the group speed error
    over the centered scheme. The regions are delimited
    by the quantity $\beta^{(n,s)}$ and the zeros of the function
    $F_{1}^{(n,s)}$ defined in (\ref{funcGroupSt}).}
  \label{RegGroupFixedOrder}
\end{figure}

Imposing  $\left\vert\hat{\epsilon}^{(n,s)}_{g}\right\vert<
\left\vert\hat{\epsilon}^{(n,0)}_{g}\right\vert$ and using
the definition (\ref{GroupErrDef}) yields
the inequality
\begin{equation}
  F^{(n,s)}_{1}(\xi) F^{(n,s)}_{2}(\xi)\left(\beta
  -\beta^{(n,s)}(\xi)\right)<0\,,
\nonumber\end{equation}
where
\begin{eqnarray}
  F^{(n,s)}_{1}(\xi)&\equiv&\partial_{\xi} f^{(n,s)}_{1}(\xi)\,,
  \nonumber\\
  F^{(n,s)}_{2}(\xi)&\equiv&\partial_{\xi} f^{(n,s)}_{2}(\xi)\,,
  \nonumber\\
  G^{(n)}(\xi)&\equiv&\partial_{\xi} g^{(n)}(\xi)\,,
  \nonumber\\
  \beta^{(n,s)}(\xi)&\equiv&\frac{G^{(n)}(\xi)}
       {F^{(n,s)}_{2}(\xi)}\,,
\label{funcGroupSt}
\end{eqnarray}
and $ f^{(n,s)}_{1,2}$ and $g^{(n)}$ are given by (\ref{funcPhaseSt}).
It is easy to see that $G^{(n)}(\xi)=-G^{(n)}(-\xi)$. 
However the signs of $F^{(n,s)}_{1,2}(\xi)$ are more difficult 
to determine.
As in the case of phase speeds analysis, we determine graphically 
(see Figure \ref{RegGroupFixedOrder}) the regions in $(\xi,\beta)$ plane where 
at fixed order of approximation, 
off-centering by $s$ points improves the accuracy of the group speed.
We see the same qualitative behavior as for the phase speeds, 
in the sense that  for sufficiently small $\beta$,
the ``+'' (``-'') speed has smaller error 
compared with the case of CFDO at least at small frequencies, 
and off-centering decreases the extent of these regions 
in $(\xi,\beta)$-space.\\
In case $s=1$, the relations (\ref{funcGroupSt}) become
\begin{eqnarray}
  F^{(n,1)}_{1}(\xi)&=& \frac{(n+1)\left\vert c_{n-1}\right\vert}{2}
  \left(\frac{n}{n+1}+\cos\xi\right)\,
  \hat{\Omega}^{2 n}\nonumber\\ 
  F^{(n,1)}_{2}(\xi)&=& \frac{(n+1)\left\vert c_{n-1}\right\vert}{2}
  \left(-\frac{n}{n+1}+\cos\xi\right)
  \hat{\Omega}^{2 n}\,,\nonumber\\  
  \beta^{(n,1)}(\xi)&=&\frac{2\left(\sign{\xi}
  -\hat{d}^{(1,n)}/\sqrt{\hat{d}^{(2,n)}}\right)}{F^{(n,1)}_{2}(\xi)}\,.
\end{eqnarray}
By analyzing the monotony of these functions using the properties 
from \ref{sectFourier},  the following result can be formulated:

At a given order of approximation $2n$, 
for sufficiently small $\beta<\frac{n}{n+1}$, 
the ``+'' group speed has smaller error 
in the case when we advect one point than in the case
when we use CFDO for all frequencies $0<\xi<\pi-\arccos\frac{n}{n+1}$,
(in the case of phase speed this was the whole range $(0,\pi)$!),
but the ``-'' speed will have larger error, at least 
for small and mid frequencies.

If $\beta>\frac{n}{n+1}$ then for both $\pm$ speeds,
in the regime of small frequencies,
the CFDO give less error than one-point advected 
scheme, while for mid and high frequencies,
the situation reverses.
The interval of small frequencies where CFDO are better
than advected scheme narrows with increasing the order
of approximation.
\subsubsection{Centered versus One-Point Upwinded Scheme, Numerically}
\begin{figure}[hbpt]
  \centering
    \includegraphics[width=0.325\textwidth,height=0.18\textheight]
		    {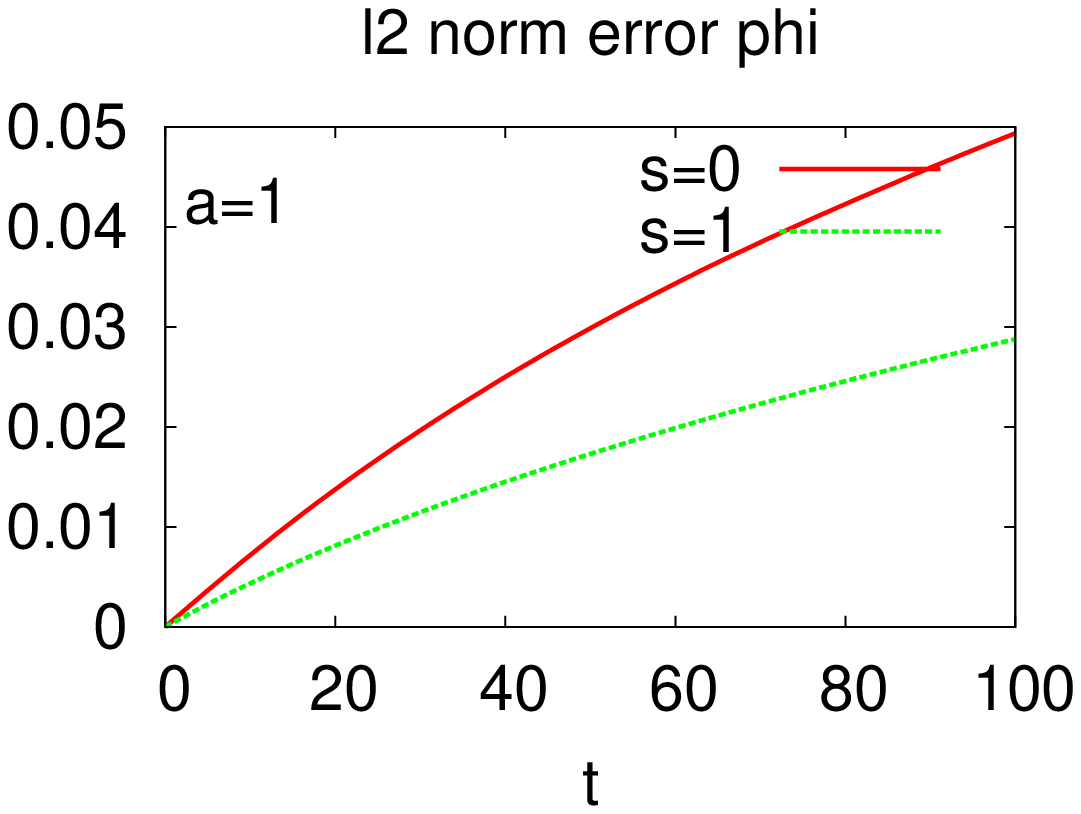}
    \includegraphics[width=0.325\textwidth,height=0.18\textheight]
		    {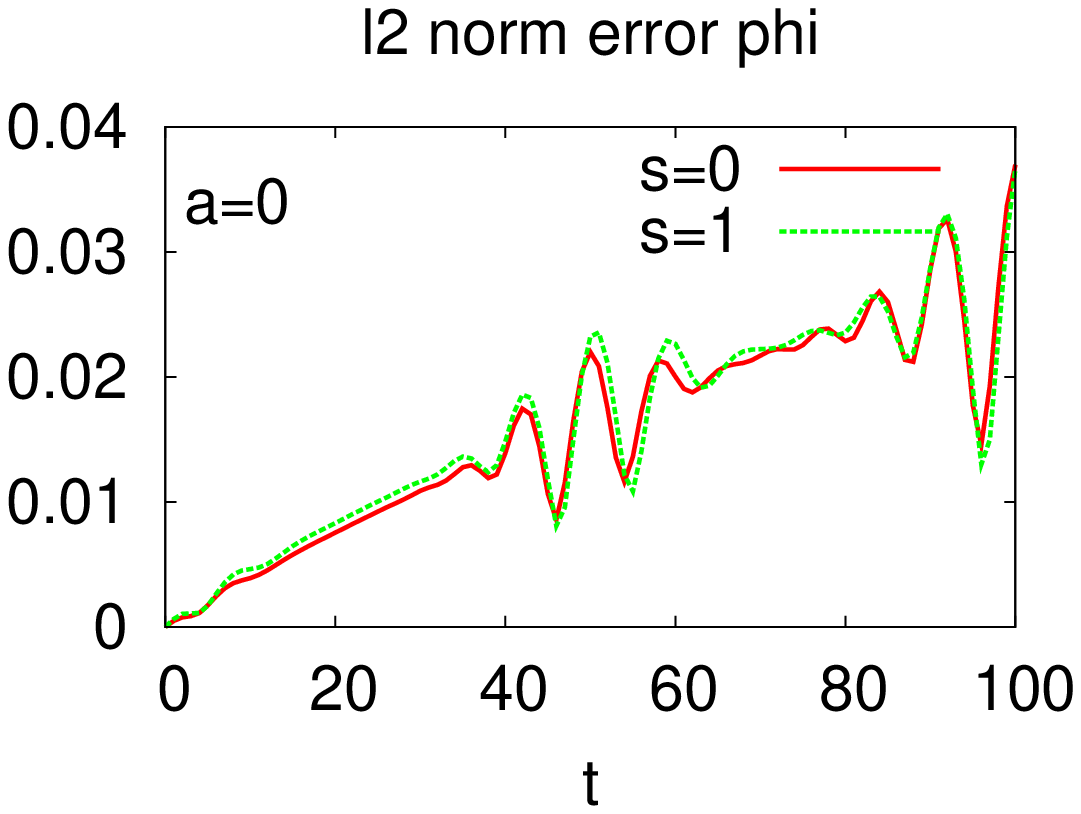}
    \includegraphics[width=0.325\textwidth,height=0.18\textheight]
		    {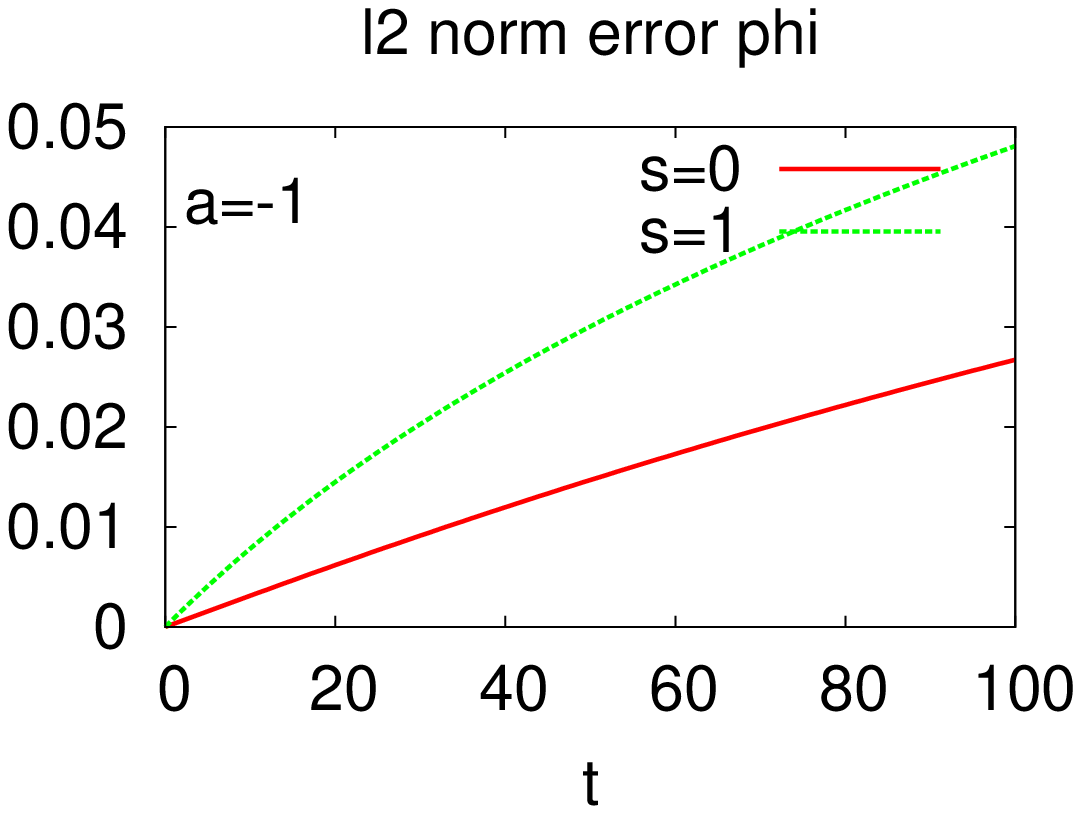}
    \includegraphics[width=0.325\textwidth,height=0.18\textheight]
		    {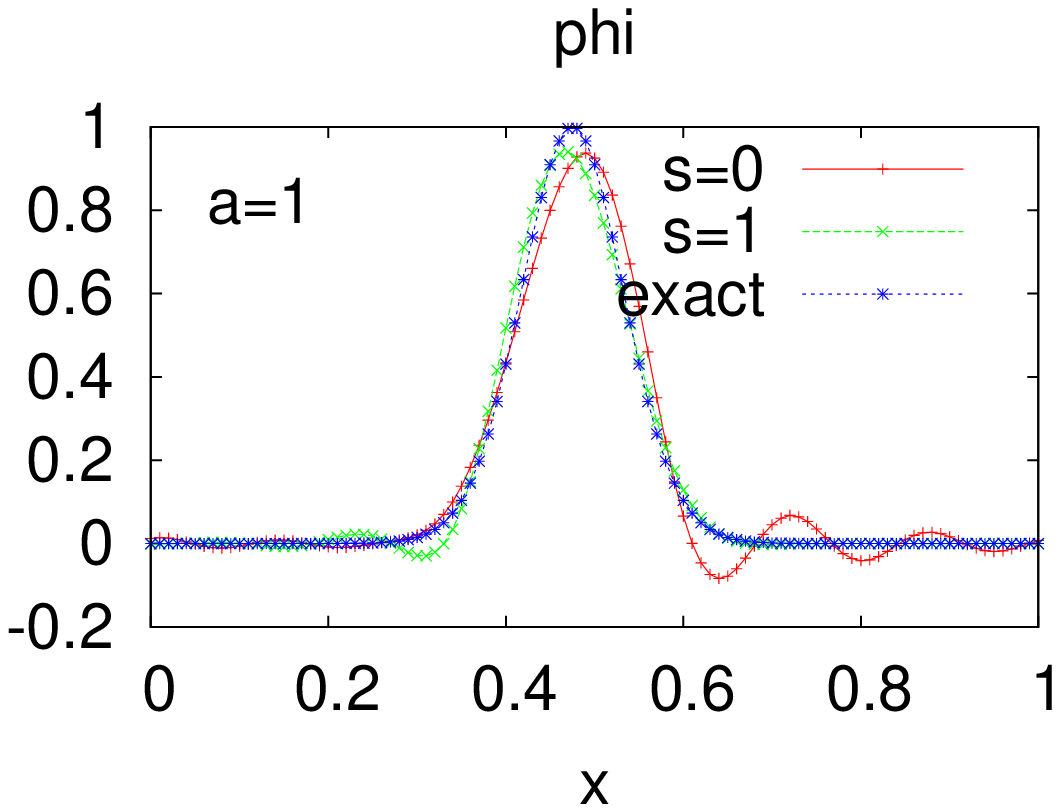}
    \includegraphics[width=0.325\textwidth,height=0.18\textheight]
		    {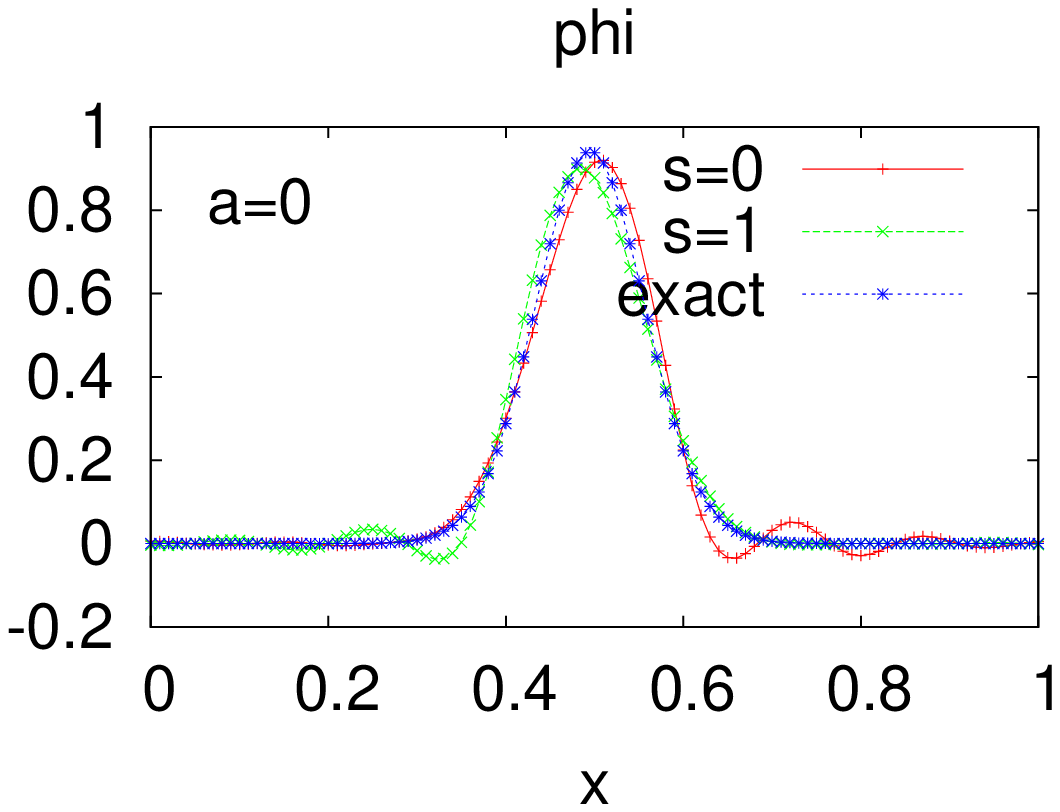}
    \includegraphics[width=0.325\textwidth,height=0.18\textheight]
		    {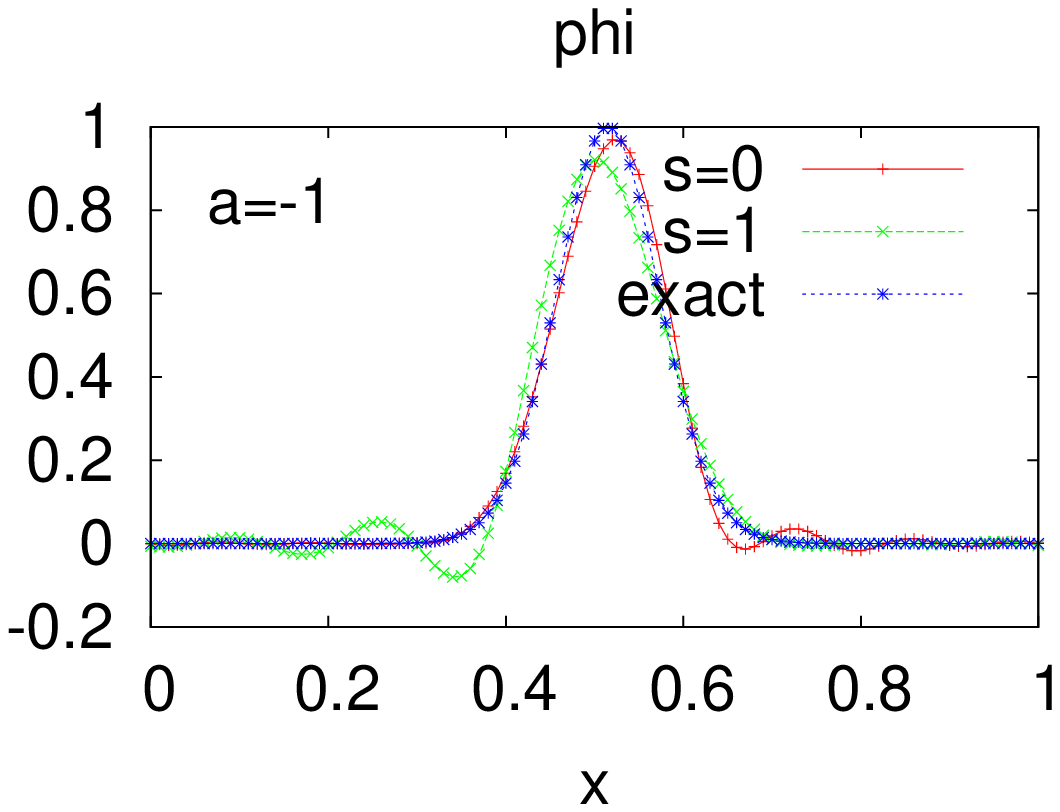}
		    \caption{The top panel shows the errors in the $l^2$ norm, 
		      the bottom panel  snapshots of $\Phi$ at $t=99 CT$.
		      The three cases from left to right are a purely
		      left-going signal ($a=1$),
                      a signal with equal amplitudes of left and right
                      going modes ($a=0$), and
                      a purely right-going signal ($a=-1$).
                      Red lines mark the centered scheme, green the  one-point
                      advected stencil, blue the exact solution.
                      For $a=1$ upwinding is more accurate, but
		      the situation reverses if the signal is
		      right-going, for $a=0$ both schemes yield
                      similar accuracy.
		    }
		    \label{NumTestSpeeds}
\end{figure}

In \ref{SpeedsFixedOrder} we showed that when 
$0<\beta\leq\frac{n}{n+1}$ the numerical ``+'' speeds
are better approximated with one-point off-centered
schemes than with centered schemes at least up to very 
high frequencies in the grid.
In this section we show some simple numerical tests 
to illustrate this fact.
We chose  $l$-periodic initial data: 
\begin{eqnarray}
  \Phi(0,x)&=&e^{-(2\pi l\tau^{2})^{-1} 
    \sin ^2\left(\frac{\pi}{l}x-\frac{\pi}{2}\right)
    }\,,
  \nonumber\\
  K(0,x)&=&a\partial_{x}\Phi(0,x),\quad x\in[0,l)\,.
\end{eqnarray}
The parameter $a\in[-1,1]$ sets
the amplitude of the ``$\pm$'' components,
\begin{equation}
  C_{\pm}=(a\pm 1)\partial_{x}\mathsf{\Phi}\,.
\nonumber\end{equation}
When $a=1(-1)$ the signal is purely ``left'' (``right'') going
and when $a=0$, the signal is equally distributed between both
modes.

We choose a grid with $N=101$ points and resolution $h=0.01$,
the width of the grid is $l=Nh=1.01$.
Also we chose $\tau=0.1$, $\beta=0.5$ 
and we integrate the wave equation
using fourth order FDOs for space derivatives
and the fourth order Runge-Kutta as time integrator.

We let $a\in\{1,0,-1\}$ and for each value of $a$
we look at the errors for the main variables 
when $s=0,1$ (see Figure~\ref{NumTestSpeeds}).
The numerical results show that, indeed, when the signal
is ``left'' going, the upwinded
scheme has less error than the centered scheme, while
when the signal is going ``right'',
the centered scheme is to be preferred.

\section{Conclusions}\label{sec:summary}
In this paper we have investigated several aspects related 
to the discretization of the  initial value problem 
for first order in time and second order in space systems of differential 
equations, using high order finite difference operators.
Special attention has been paid to the situation 
when some of the first derivatives are approximated 
with off-centered discrete operators as is customary 
for treating black hole spacetimes in numerical relativity.
Our investigation has been divided into three parts: 
(a) We started with an analysis of certain 
properties of the finite difference operators (Section \ref{sec:FDO}).
(b) Using these properties we have extended 
the validity of an existing stability method (Section \ref{sec:StabMethod}).
(c) We analyzed the stability and the numerical speeds 
in the case of the scalar wave equation (Section \ref{sectionWave}). 

In the following we will give a brief overview of the results.
\subsubsection*{(a) Analysis of first and second order 
discrete derivative operators}
A set of mathematical properties have been deduced for the Fourier symbols 
associated with the second order centered and first order 
(not necessarily centered) discrete derivatives. They 
are in the form of inequalities, recursive  and 
differential relations for the Fourier symbols 
at different orders of approximation or at different off-centerings.
Here we mention two of them:
 
While first derivatives do not converge in the limit 
$n\rightarrow\infty$ at the maximum
grid frequency ($\xi = \pi$), second derivatives do converge
at all frequencies (that is the highest frequency in the grid will
not be captured by the first order derivative, regardless of 
the order of approximation or the off-centering,
while the second centered derivative can ``see'' it and 
approximates it better with increasing order).

For first order derivatives, 
increasing the off-centering ($s$) at a fixed order of 
approximation increases the error of 
the derivative at small frequencies.
At larger frequencies this behavior changes.
E.g., for $s=1$, beyond a certain frequency $\xi^{(n)}$,   
the error is smaller than for the case $s=0$.
As a consequence, off-centering of the first order derivative 
in the case of the advection equation, increases the error at small 
frequencies, while at high frequencies, this situation can reverse.

\subsubsection*{(b) Generalization of a stability analysis method}
In \cite{CalHinHusa}, necessary and sufficient conditions for stability 
have been deduced assuming that (\ref{contsystem}) is discretized 
using  2nd or 4th order CFDOs and  integrated in time using a time integrator
locally stable on the imaginary axis.
The validity of this stability method is extended here to 
$2n$-order spatial accuracy,
including also the case where some derivatives are 
approximated with non-centered FDOs 
%
and dissipation is added to the system. 
It is pointed out that neither adding artificial dissipation 
nor shift advection terms 
affects the eigenvectors of the discrete symmetrizer,
and thus the conditions for semi-discrete numerical stability.  
The Courant limit will of course be affected in general.

\subsubsection*{(c) Application: Scalar Shifted Wave Equation}
The stability method presented in Section \ref{sec:StabMethod} is applied  
to the case of the wave equation on a curved background in $d+1$ dimensions. 

In the case of  1-D shifted wave equation in flat spacetime, the 
Courant limits and the numerical speeds have been analyzed in detail 
in respect to the order of approximation and off-centering of the first 
derivative.
\begin{itemize}
\item  \textbf{Courant limits}
\end{itemize} 
  Off-centerings by more than one point require dissipation for stability.
  In these cases, the minimal Kreiss-Oliger dissipation 
  needed for stability has been computed and found 
  to be proportional to the shift $\beta$.
  For centered schemes, higher order approximations have
  lower Courant limits. Interestingly, 
  this does not hold  for off-centered schemes
  (when adding just dissipation to be in the
  local stability regime) --- for large enough shift, the Courant limit
  is actually larger for higher order schemes.
  Off-centering generally reduces the Courant limit drastically, except 
  for at least fourth order accurate schemes, when only
  one-point off-centering is used: for higher than fourth order schemes 
  one-point off-centering only leads to a minor reduction of the CFL
  factor.
  \begin{itemize} 
  \item \textbf{Numerical speeds} 
  \end{itemize} 
  Without shift, higher order approximations always
  result in more accurate numerical speeds, 
  with nonzero shift this is not generally true at higher frequencies.
  
  Although the truncation 
  error for the first order derivative increases with the 
  off-centering, the mixing with the
  second order discrete derivative in the scheme,
  causes upwinded stencils to give a higher overall accuracy in some
  situations.
  
  More precisely,  it is shown that advecting
  shift terms by an odd (even) number of points
  reduces the errors of the ``$+$'' (``$-$'') numerical speeds
  in some intervals of the spectrum that 
  include the small frequency range, if the shift is not too large.
  The extent of the regions in the (frequency, shift)-parameter space
  where this improvement appears,  decreases with off-centering, 
  in such a way that for $s=1$ one gets the strongest
  effect. 

  Thus, at a given order $2n$, if the shift satisfies 
  $0<\beta<\frac{n}{n+1}$,  
  then off-centering by one point 
  has in comparison with the centered scheme, 
  better ``$+$'' phase speed error for all frequencies,
  and better ``$+$'' group speed error  for all frequencies
  up to a very high frequency in the grid,
  $\pi-\arccos\frac{n}{n+1}$. 
    
  If the wave equation is written 
  in first order form and discretized using CFDO, then 
  for a given order of approximation, the second order system 
  discretized also with CFDO has smaller phase and group errors 
  than the first order one, 
  if $\left\vert\beta\right\vert\leq\frac{2n+3}{4(n+1)}$.
  If $\left\vert\beta\right\vert$ is not in this interval then 
  one pair of speeds (phase and group) is better approximated 
  by the second order system,
  while the other one is better approximated
  by the first order system.

  A detailed understanding of finite difference algorithms for first order in time, second order
  in space systems, in particular as applied to the Einstein equations, will require significant
  further work. Already for the shifted wave equation, it will be interesting to study the
  errors in the multidimensional case, e.g.~when the wave propagates in a direction that
  is not aligned with the grid. A similar analysis for the full 
  Einstein equations will require a substantial use of computer algebra methods. We also point out
  that for the Einstein equations much of the complications come from the nonlinear source terms,
which are beyond the scope of our present analysis.

\subsection*{Acknowledgments}
M. Chirvasa would like to thank Bela Szilagyi for inspiring 
discussions at the early stages of this work, and we thank Jeffrey Winicour
and Gerhard Zumbusch for helpful comments on our manuscript.
S. Husa has been in supported in part as a VESF fellow of the European Gravitational Observatory
(EGO), by DFG grant SFB/Transregio~7 ``Gravitational Wave Astronomy'', by DAAD grant D/07/13385  
and grant FPA-2007-60220 from the Spanish 
Ministerio de Educaci\'on y Ciencia.

\begin{appendix}

\section{Explicit Expressions for Finite Difference Operators}\label{constr_prop_1D_explicit}

  Explicit formulas for general finite difference operators in one
  dimension can be constructed in a surprisingly simple way by use of the
  in the following lemma.
  \begin{lemma}
    Consider a general finite difference operator 
  $D^{(m,n,s,\epsilon)}$, which is written as a linear combination of shift 
  operators of the form (\ref{genformFDOexplicit})
  \begin{equation}
    D^{(m,n,s,\epsilon)}=
    h^{-m}\sum_{k=-n+\epsilon s}^{n+\epsilon s}\tilde{f}_{m,n,s,\epsilon,k}S^{k}\,.
    \nonumber
  \end{equation} 
    Then the coefficients $\tilde{f}_{m,n,s,\epsilon,k}$
    are the coefficients of $y^{k}$ in the Taylor expansion of the function 
    $$f^{m,n,s,\epsilon}(y)=y^{n-\epsilon s}(\ln y)^{m}$$ 
    around the point $y_{0}=1$ up to the order $(y-y_{0})^{2n}$.
    In general, the accuracy of this operator will be $2n+1-m$.
  \end{lemma}
  To prove the lemma, it is enough to 
  consider the scalar function $\mathsf{v}(x)=e^{i\omega x}$ with 
  $\omega\in \Complex$ and the associated grid function.
  By applying accordingly the differential and discrete operators 
  we get:
  \begin{eqnarray}
    \partial^{m}\mathsf{v}(x_{\nu })&=&(i\omega)^{m}e^{i\omega h \nu }
    \nonumber\\
    D^{(m,n,s,\epsilon)}v_{\nu }&=& h^{-m}\sum_{k=-n+\epsilon s}^{n+\epsilon s}
    \tilde{f}_{m,n,s,\epsilon,k}e^{i\omega h(\nu +k)}
    \nonumber\\
    \partial^{m}\mathsf{v}(x_{\nu })&=&D^{(m,n,s,\epsilon)}v_{\nu }+\mathcal{O}(h^{q})
    \label{relabc}
  \end{eqnarray}
  Introduce $y=e^{i\omega h}$ which gives $i\omega=h^{-1}\ln y$.
  The relations (\ref{relabc}) lead to
  \begin{equation}
    y^{n-\epsilon s}(\ln y)^{m}= \sum_{k=-n+\epsilon s}^{n+\epsilon s}
    \tilde{f}_{m,n,s,\epsilon,k}y^{k+n-\epsilon s}
    +\mathcal{O}(h^{q+m})y^{-\nu +n-\epsilon s}\,.
    \nonumber
  \end{equation}
  In the limit $h\rightarrow 0$, $y\rightarrow 1$.
  The function $y^{n-\epsilon s}(\ln y)^{m}$ is now Taylor expanded 
  around the point $y_{0}=1$ up to $(y-y_{0})^{2n}$ and 
  the coefficients $\tilde{f}_{m,n,s,\epsilon,k}$ are identified.
  What remains is:
  \begin{equation}
    \mathcal{O}((y-1)^{2n+1}) = \mathcal{O}(h^{q+m})y^{-\nu +n-\epsilon s}\,.
    \nonumber
  \end{equation}
  After replacing $y$  with its definition and taking the limit $h\rightarrow 0$
  we obtain $q=2n+1-m$ and the lemma is proved.
  \begin{corollary}
    The FDOs associated with the first and second derivative are 
  \end{corollary}
  \begin{eqnarray}
    D^{(1,n,s,\epsilon)}&=&h^{-1}
    \sum_{k =-n+\epsilon s}^{n+\epsilon s}\alpha_{n,s,\epsilon,k }S^{k },
    \nonumber
    \\
    D^{(1,n)} \equiv D^{(1,n,0,0)} &=&h^{-1}\sum_{k =1}^{n}
    \frac{k \beta_{n,k }}{2}\left(S^{k }-S^{-k }\right),
    \nonumber
    \\
    D^{(2,n)} \equiv D^{(2,n,0,0)} &=&h^{-2}\sum_{k =0}^{n}
    \beta_{n,k }\left(S^{k }+S^{-k }\right),
 \label{explicitFDO}
  \end{eqnarray}
  where 
  \begin{equation}
    \alpha_{n,s,\epsilon,k } =
    \left\{
    \begin{array}{cc}
      \frac{(-1)^{k +1} (n+s)! (n-s)!}{k  (n+\epsilon s-k )! (n-\epsilon s+k )!},&k \neq 0\\\\
      \epsilon\left(H_{n-s}-H_{n+s}\right),&k =0\\
    \end{array}
    \right.
    \quad\text{and}\quad
    \beta_{n,k }=
    \left\{
    \begin{array}{cc}
      \frac{2(-1)^{k +1}\left(n!\right)^2}
      {k ^2\left(n+k \right)!\left(n-k \right)!} &k \geq 1\\\\
      -\sum_{k =1}^{n}\beta_{n,k } &k =0 \, .\\
    \end{array}
    \right.   \nonumber
  \end{equation}
  In the relations above,  $H_{n}=\sum_{j=1}^{n}j^{-1}$ 
  is the harmonic number.
  Note that $k \beta_{n,k }=2\alpha_{n,0,0,k }$ for $k \geq 1$.

\section{Finite difference operators in $d$-dimensions}
\label{sec:GenTo_dDim}
Consider a $d$-dimensional grid defined by the set of points 
 $x_{\nu}=(\nu_{1}h_{1},\dots,\nu_{d}h_{d})$, 
 where  $\nu=(\nu_{1},\dots, \nu_{d})$ is a multiple index, 
 $\nu_{j}\in\Integer$
 and $h_{j}$ represents the grid spacing in $j$-direction 
 ($j=1,\dots,d$).
 Corresponding to the continuum vector-function 
  $\mathsf{v}:\Real^{d}\rightarrow \Complex\times\dots\times\Complex$
 we associate the grid vector-function $v$ such that 
 $v_{(\nu)}\equiv v(x_{(\nu)})=\mathsf{v}(x_{(\nu)})$.
 
 The shift operator by $k$-points in the $j$-direction,  $S_{j}^{k}$ 
 is defined by
 \begin{equation}\label{eq:shiftop_d-dim}
    S_{j}^{k}v_{(\nu_{1},\dots, \nu_{d})}=v_{(\nu_{1},\dots \nu_{j}+k,\dots \nu_{d})}\,.
  \end{equation}
  A discrete operator $D_{j}$ acting in the $j$-direction is constructed 
  as a linear combination of the shift operators defined in 
  (\ref{eq:shiftop_d-dim})  using the same weights as 
  the corresponding one dimensional operator $D$; 
  an operator  $D_{j_{1}\dots j_{r}}$   acting in $j_{1}\dots j_{r}$-directions 
  is constructed as a composition of one-directional operators 
  $\{D^{1}_{j_{1}},\dots,D^{r}_{j_{r}}\}$:
  \begin{eqnarray}
    D_{j}&=&\sum_{k}a^{k}(h_{j})S^{k}_{j}\quad\text{where}\quad
    \quad D=\sum_{k}a^{k}(h)S^{k}
    \label{constrOneDirOP}\\
    D_{j_{1}\dots j_{r}}&=&
    a(h_{j_{1}},\dots h_{j_{r}})D^{1}_{j_{1}}\dots D^{r}_{j_{r}}
    \label{constrMoreDirOP}
\end{eqnarray}
In order to represent the functions in Fourier space, 
we consider only grid function which are periodic in each direction 
and limit the grid to having a finite number of points,   
$N_{j}$ for the direction $j$, $j=1,\dots,d$. 
We introduce 
\begin{eqnarray}
  h&=&(h_{1},\dots,h_{d}),\quad\quad N=(N_{1},\dots,N_{d}),\nonumber\\
  b_{\nu}(\omega)&=&(2\pi)^{-d/2}e^{i\omega x_{\nu}},\nonumber\\
  V_{h}&=&h_{1}\dots h_{d},\nonumber\\
  \mathcal{S}_{\underline{x}}(N)
  &=&\mathcal{S}_{\underline{x}}(N_{1})\times\dots\times
  \mathcal{S}_{\underline{x}}(N_{d}),\nonumber
  \\
  \mathcal{S}_{\underline{\omega}}(N) &=&
    \mathcal{S}_{\underline{\omega}}(N_{1})
    \times\dots\times \mathcal{S}_{\underline{\omega}}(N_{d}),\nonumber
    \\
   \mathcal{S}_{\underline{\xi}}(N) &=&
   \mathcal{S}_{\underline{\xi}}(N_{1})
   \times\dots\times \mathcal{S}_{\underline{\xi}}(N_{d})\,.
    \nonumber
  \end{eqnarray}
  In the relations above, 
  $ \mathcal{S}_{\underline{x}}(N_{j}\in\Natural)$, 
  $\mathcal{S}_{\underline{\omega}}(N_{j}\in\Natural))$ 
  and 
  $\mathcal{S}_{\underline{\xi}}(N_{j}\in\Natural)$
  have been defined in (\ref{defSgridpoints1D}), (\ref{defSomega1D}) 
  and (\ref{defSxi1D}), respectively. 
  Then, the formulas for the Fourier decomposition 
  (\ref{FourierRepresGridFunc1D}), scalar product 
  and a norm (\ref{normh1d}) and Parseval relation  
  (\ref{Parseval1d}) are valid also in $d$-dimensions.

  Let $\xi\in\mathcal{S}_{\underline{\xi}}(N)$
  and apply the shift operator by $k$-points in the $j$-direction 
  $S^{k}_{j}$ on a   basis vector $ b_{\nu}(\omega)$.
  This leads to 
  \begin{equation}
    S^{k}_{j}b_{\nu}(\omega)
    =\hat{S}^{k}(\xi_{j})
    b_{\nu}(\omega),
    \quad \text{with}\quad
    \hat{S}^{k}(\xi)=e^{i\xi k}\,.
    \nonumber
  \end{equation}
  Then the Fourier symbols for the operators $D_{j}$ and $D_{j_{1}\dots j_{r}}$ 
  from (\ref{constrOneDirOP}-\ref{constrMoreDirOP}) are defined by:
  \begin{eqnarray}
    D_{j} b_{\nu}(\omega)
    &=&\hat{D}(\xi_{j};h_{j})b_{\nu}(\omega)\nonumber\\
    D_{i_{1}\dots i_{r}}b_{\nu}(\omega)
    &=&\hat{D}(\xi_{i_{1}},\dots \xi_{i_{r}};h_{i_{1}},\dots h_{i_{r}})
    b_{\nu}(\omega)\nonumber\\
    \text{with}&&\nonumber\\
    \hat{D}(\xi_{j};h_{j})&=&\sum_{k}a^{k}(h_{j})\hat{S}^{k}(\xi_{j}),
    \nonumber\\
    \hat{D}(\xi_{i_{1}},\dots \xi_{i_{r}};h_{i_{1}},\dots h_{i_{r}})&=&
    a(h_{i_{1}},\dots h_{i_{r}})\hat{D}^{1}(\xi_{i_{1}})\dots \hat{D}^{r}(\xi_{i_{r}})
    \nonumber
  \end{eqnarray}
\section{Further Properties of the Fourier symbols}
\label{appPropFourier}
\begin{enumerate}
\item \textbf{Recurrence relations}:
  \begin{eqnarray*}
    \hat{d}^{(1,n+1)}&=&\hat{d}^{(1,n)}
    +\hat{\delta}\left\vert c_{n}\right\vert
    \hat{\Omega}^{2n}\,.\\
      \hat{d}^{(2,n+1)}&=&\hat{d}^{(2,n)}
      +\left\vert d_{n}\right\vert
      \hat{\Omega}^{2n+2}\,.\\
      \hat{d}^{(1,n,s)}&=&
      \hat{d}^{(1,n,s-1)}
      -\frac{(-1)^s 
      }{(n+s)C_{2n}^{n+s}}
      \hat{\Omega}^{2 n+1}
      \cos \frac{\left(2s-1\right)\xi}{2}
      \\
	\hat{\mathbf{d}}^{(1,n,s)}
      &=&\hat{\mathbf{d}}^{(1,n,s-1)}
	-\frac{(-1)^s 
	}{(n+s)C_{2n}^{n+s}}
      \hat{\Omega}^{2 n+1}
	\sin 
	\frac{\left(2s-1\right)\xi}{2}
    \end{eqnarray*}
\item \textbf{Small frequency behavior}:
  \begin{eqnarray*}
    \hat{d}^{(1,n,s)}&\simeq&\xi\left[1
    -(-1)^{n}T^{(1,n,s)}\xi^{2n}\right]
         \\
    \hat{\mathbf{d}}^{(1,n,s)}&\simeq&
    (-1)^{n}s\frac{2n+1}{2n+2}T^{(1,n,s)}\xi^{2n+2}
        \\
    \sqrt{\hat{d}^{(2,n)}}&\simeq&
    \xi\left[1
      -\frac{(-1)^{n}}{2}T^{(2,n)}\xi^{2n}\right]
   \end {eqnarray*}
\item \textbf{Sums}
  \begin{eqnarray*}
       \sum_{k=0}^{\infty}
       \left|c_{k}\right|x^{2k}&=&\frac{\arcsin\frac{x}{2}}{\frac{x}{2}
	 \sqrt{1-\left(\frac{x}{2}\right)^2}}\,,\\
       \sum_{k=0}^{\infty}
       \left|d_{k}\right|x^{2k}&=& 
       \left(\frac{\arcsin\frac{x}{2}}{\frac{x}{2}}\right)^2\,.
  \end{eqnarray*}
\item \textbf{Limits $n\rightarrow \infty$}:
\begin{eqnarray*}
    \underset{n\rightarrow\infty}{\lim}\hat{d}^{(2,n)}(\xi)
  &=&\xi^{2}\,,\quad \forall \xi\in[0,\pi]\,,
    \\ 
  \underset{n\rightarrow\infty}{\lim}\hat{d}^{(1,n,s)}(\xi)
  &=&\xi\,,\quad \forall \xi\in[0,\pi)\,,
    \\
  \underset{n\rightarrow\infty}{\lim}\hat{\mathbf{d}}^{(1,n,s)}(\xi)
  &=&0\,,\quad \forall \xi\in[0,\pi]\,,
  \\
  \underset{n\rightarrow\infty}{\lim}\hat{r}^{(n)}(\xi)&=&0,
  \quad \forall \xi\in[0,\pi)\,.
\end{eqnarray*}
\item \textbf{The $D^{(2,n)}$-norm} defined by
  \begin{equation*}
    \left\Vert v \right\Vert^{2}_{h,D^{(2,n)}}=\frac{1}{h^{2}}
    \sum_{i=1}^{d}\sum_{k=1}^{n}\left|d_{k-1}\right|
    \left\Vert (hD_{+i})^{k}u\right\Vert_{h}^{2}
    +
    \left\Vert v\right\Vert^{2}_{h}\,.
  \end{equation*}
  is equivalent with the $D_{+}$ norm.
  This norm has been used to prove strong stability of the initial boundary 
  value problem for the wave equation in \cite{GundCal}
  for the second and fourth order accuracy case.
\end{enumerate}
\end{appendix}

\end{document}